\documentstyle[html,aps,twocolumn,epsfig,tighten,floats]{revtex}

\newcommand{\eq}[1]{{\frenchspacing Eq.~(\ref{#1})}}
\newcommand{\fig}[1]{{\frenchspacing Fig.~(\ref{#1})}}

\newcommand{\trq}{\mbox{Tr}\;Q}
\newcommand{\beq}{\begin{equation}}
\newcommand{\eeq}{\end{equation}}

\begin{document}

\title{On the low fermionic eigenmode dominance\\
 in QCD on the lattice}

\author{H.\ Neff$^{1}$\thanks{Electronic mail: neff@theorie.physik.uni-wuppertal.de},
        N.\ Eicker$^{2}$,
        Th.\ Lippert$^{2}$, 
        J.\ W.\ Negele$^{3}$, and
        K.\ Schilling$^{1,2}$} 
\address{$^{1}$John von Neumann Institute for Computing Research Center J\"ulich, 52425
        J\"ulich, Germany\\
         $^{2}$Dept.\ of Physics, University of Wuppertal, 42097
        Wuppertal, Germany\\
         $^{3}$Center for Theoretical Physics, MIT, Cambridge, USA}

\maketitle

\begin{abstract}
  We demonstrate the utility of a spectral approximation to fermion
  loop operators using low-lying eigenmodes of the hermitian
  Dirac-Wilson matrix, $Q=\gamma_5 M$.  The investigation is based on
  a total of 400 full QCD vacuum configurations, with two degenerate
  flavors of dynamical Wilson fermions at $\beta =5.6$, at two
  different sea quark masses. The spectral approach is highly
  competitive for accessing both topological charge and disconnected
  diagrams, on large lattices and small quark masses.  We propose
  suitable partial summation techniques that provide sufficient
  saturation for estimating $\trq ^{-1}$, which is related to the
  topological charge. In the effective mass plot of the $\eta'$ meson
  we achieved a consistent early plateau formation, by ground state
  projecting the connected piece of its propagator.
\end{abstract}

\section{introduction}
The Euclidean approach of lattice gauge theory (LGT) has been established as a
viable framework to deal with quantum chromodynamics beyond the limitations of
perturbation theory~\cite{Ecfa:2000rp}.  Whereas practical algorithms have been
developed to calculate many physical observables, there is a class of physical
quantities that is highly sensitive to fluctuations in the vacuum gauge field
that has not been amenable to known {\it ab initio} numerical methods.
Notorious examples of such computationally intractable problems include mixing
phenomena between fermionic and glueball states~\cite{McNeile:2000xx},
disconnected quark loop diagrams occurring in flavor singlet matrix
elements~\cite{Okawa:1996jp,Gusken:1999as}, and the infamous $\eta'$
propagator~\cite{AliKhan:1999zi,Struckmann:2000bt,McNeile:2000hf}.

In all these instances one would like to take advantage of
self-averaging effects by exploiting the translational invariance of
the QCD ground state.  This amounts to probing with objects shifted
across all lattice space-time points and hence to the computation of
light quark propagators on the entire vacuum field, i.e.\ the full
inverse of the Dirac operator, $M$, which exhibits a high condition
number in the regime of light quark masses.  Unfortunately this is a
prohibitively expensive numerical task, with $M$ being of rank $ >
10^6$ on typical lattice sizes. Therefore the standard approach is to
resort to stochastic estimator
techniques~\cite{Bitar:1989bb,Dong:1995rx,Eicker:1996gk,Michael:1998rk,Wilcox:1999ab,Viehoff:1997wi,Michael:1999nq}.

The approach to these previously intractable problems explored in this work is
motivated by the expectation that low-lying modes of the Dirac operator should
embody the important features of fermionic physics in the chiral
regime~\cite{Ivanenko:1998nb}~\footnote{Early pioneering work in this
  direction has been done by Barbour et
  al.~\cite{Barbour:1984ud,Setoodeh:1988ds}.}.  Physically, topological
excitations in the QCD vacuum, corresponding to instantons in the
semiclassical limit, generate low-lying fermion modes that play an essential
r\^ole in light quark physics, ranging from generating the 't Hooft interaction
to producing the chiral condensate.  Hence, for sufficiently light quark
masses, expansion in a basis that contains these low-lying eigenmodes should
cover the essential physics associated with observables involving the
topological charge, flavor singlet disconnected diagrams, and processes
described by the 't Hooft interaction. 

It therefore appears worthwhile to
launch another attempt to explore the potential of spectral
methods/approximations~\cite{Venkataraman:1997xi}. The question is, how many
low-lying eigenmodes suffice to bear out the important features of long range
physics in practical instances.

There are two options to proceed, based on the spectral representations of the
non-normal Wilson-Dirac matrix $M$ and the hermitian matrix $Q=\gamma_5 M$.
The advantage of $M$ lies in its shiftable (with respect to $\kappa$)
structure $M = 1 - \kappa D$, but it requires to work with biorthogonal sets
of eigenmodes and complex eigenvalues. The hermiticity of $Q$, on the other
hand, allows for a simple ordering of the orthogonal eigenmodes and a natural
definition of low-lying eigenmodes.

We decided to work with $Q:=\gamma_5 M$ and its
 eigenmodes~\cite{Jansen:1997iz,Narayanan:1998iq,Duncan:1999nx} \beq
 Q | \psi_i \rangle = \lambda_i | \psi_i \rangle \; . \eeq We can
 express the quark propagator in terms of the `eigenmodes`
 $(\lambda_i,| \psi_i \rangle)$ from the spectral representation:
\begin{eqnarray}
&Q^{-1}&(n,\alpha,a;m,\beta,b)
= \nonumber\\ &&\sum_{i =
1}^{V}\frac{1}{\lambda_i} \frac{ | \psi_i (n,\alpha,a) \rangle \langle
\psi_i (m,\beta,b) | } { \langle \psi_i | \psi_i \rangle}\;
,\label{eq:inverse}
\end{eqnarray}
where $n,\alpha,a$ etc.\ label lattice sites, Dirac and color indices,
respectively.

For our practical benchmarking we use 200 (195) SESAM lattices of size
$16^3 \times 32$ with $\kappa = 0.1575$
($0.1560$)~\cite{Eicker:1998sy}. Thus the rank of $Q$ is
$\mbox{dim}(Q) = 1\,572\,864$, which bars us from complete
diagonalization of $Q$ and the use of the {\it identity} as such,
\eq{eq:inverse}.  Nevertheless, the low-lying eigenvalues, i.e.\ the
ones with smallest moduli, represent a large weight and are expected
to saturate the sum in the regime of small quark masses.

In the present paper we wish to investigate to what extent the
spectral sum allows, under the conditions of the SESAM simulations,
for a truncation of the propagator at a reasonable (in the sense of
practicality) number of lowest-lying eigenmodes. As a testbed for our
truncated eigenmode approach (TEA), we will use $\trq ^{-1}$ in
addition to the pion and $\eta'$-correlators.

\begin{figure}[tp]
\centering{\epsfig{figure=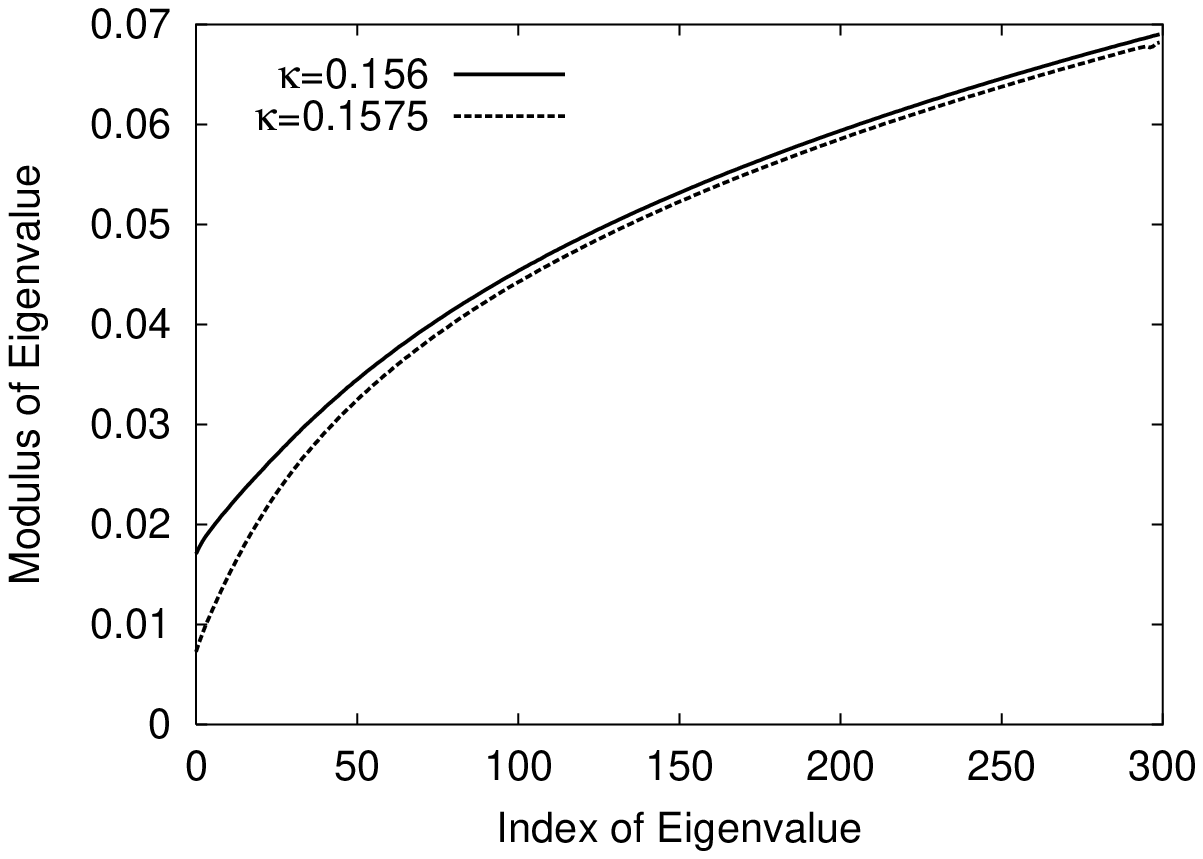,width=1.0\linewidth}} 
\vskip .5cm
\caption{Distribution of moduli of the eigenvalues of $Q$, averaged
over the configurations.  The upper line corresponds to
$\kappa=0.156$, the lower to $\kappa=0.1575$.}
\label{fig:massdep}
\centering{\epsfig{figure=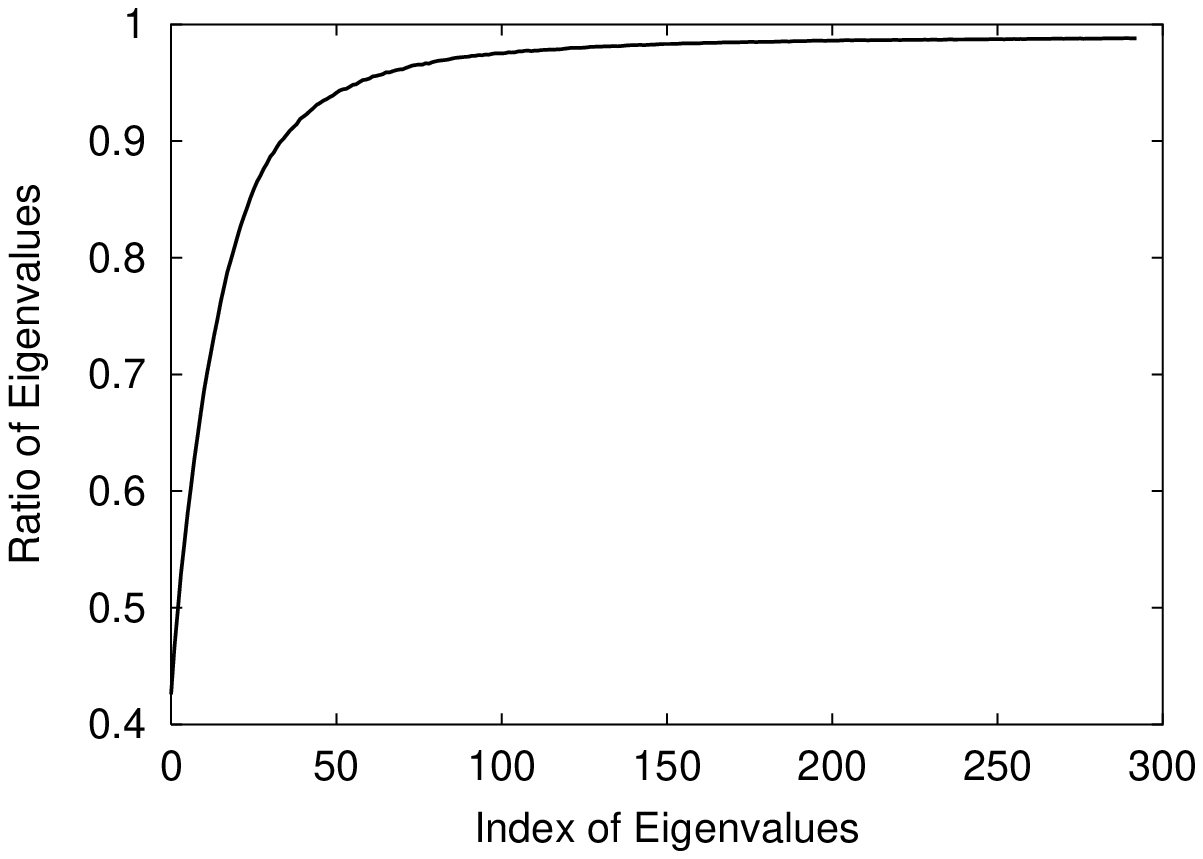,width=1.0\linewidth}} 
\vskip .5cm
\caption{Same as Fig.~\ref{fig:massdep}, but plotted in terms of eigenvalue
ratios, $|\lambda_i(0.1575)/\lambda_i(0.1560)|$, both ordered
according to Eq.~\ref{eq:nat_order}.}
\label{fig:ratio}
\end{figure}

\section{computing low-lying eigenmodes} 
\begin{figure}[t]
\centering{\epsfig{figure=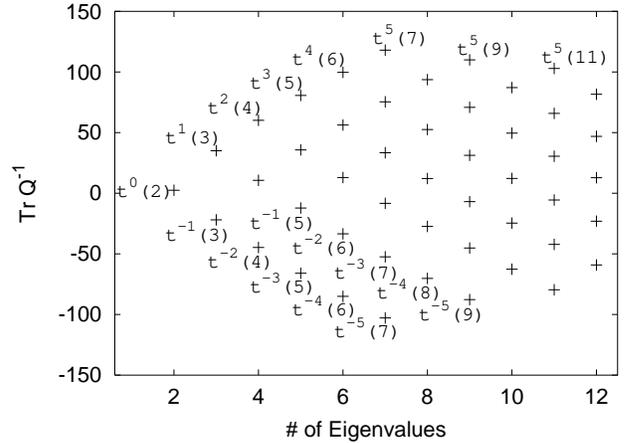,width=1.0\linewidth}} 
\vskip .5cm
\caption{Partial sums, according to Eqs.~\ref{eq:series1} and \ref{eq:series2},
from a particular SESAM configuration at $\kappa = 0.1575$.}
\label{fig:top1}
\end{figure}
\begin{figure*}[ht]
\centering{\epsfig{figure=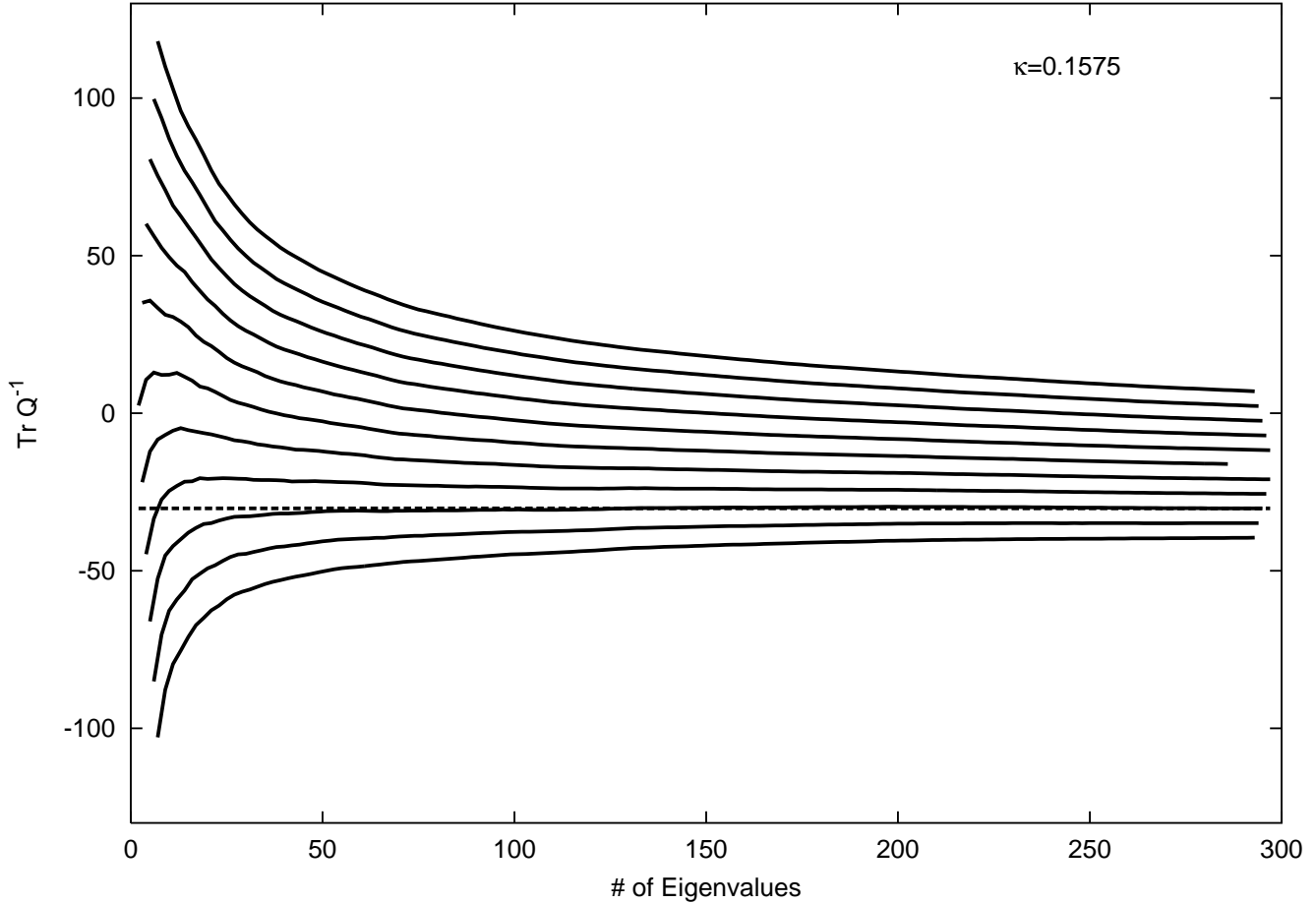,width=1.0\linewidth}} 
\vskip .5cm
\caption{Same as in Fig.~\ref{fig:top1}.
 The full lines connect the entries, $t^j$,  with equal values of $j$. The
horizontal line corresponds to the value $t$ as defined in Eq.~\ref{eq:plateau}.} 
\label{fig:top2} 
\end{figure*}

The hermitian operator $Q$ is defined through
\begin{displaymath}
Q = \gamma_5 M \; ,
\end{displaymath}
where  $M$ is the standard Wilson-Dirac matrix 
\begin{eqnarray}
&&M(n,\alpha,a;m,\beta,b) = \frac{1}{2 \kappa}
\delta(n,\alpha,a;m,\beta,b) \\ &&- \frac{1}{2} \sum_{\mu =1}^4 \Big[
\big( \delta(\alpha;\beta) - \gamma_{\mu}(\alpha;\beta) \big)
U_{s_1,\mu}(a;b) \delta(n+\hat{\mu};m) \nonumber\\ &&
\hspace{.5cm}+\big( \delta(\alpha;\beta) + \gamma_{\mu}(\alpha;\beta)
\big) U^{\dagger}_{n-\hat{\mu},\mu}(a;b) \delta(n-\hat{\mu};m) \Big]\;
. \nonumber
\label{eq:dirac} 
\end{eqnarray}
Here the $\gamma_{\mu}$ are the Dirac matrices, $U$ is the gauge field (vacuum
configuration), $\delta$ is the Kronecker delta function and $\hat{\mu}$
denotes a unit vector in a space or time direction.

The eigenmodes are determined by use of the parallelized Arnoldi
package\footnote{We employ the Arnoldi method as provided by the
parallel Arnoldi Package PARPACK~\cite{PARPACK} from Rice University.
The Arnoldi method is designed for non-hermitian matrices, but it
reduces, when applied to a hermitian matrix, to the Lanczos
method.}~\cite{PARPACK}. We speed up the computations by application
of the polynomial acceleration technique~\cite{Gattringer:1998xn}.

The Arnoldi method works efficiently when calculating eigenvalues on
the surface of the spectrum, here around $\lambda_{min}$ and
$\lambda_{max}$.  For this reason we first need to execute a
preparatory step where we map, by means of a suitable polynomial $q: Q
\mapsto q(Q)$, the low-lying modes from the inside of the spectrum to
the surface.  We will contrive the polynomial such that the spectrum
is already prepared for the subsequent Chebyshev acceleration
step~\cite{Radke:1996ri}.

\paragraph{First step: Mapping}
We apply a very simple polynomial transformation $q$. To be specific
 we proceed as follows:
\begin{itemize}
\item[-] Target a spectral window of $Q$ within which all eigenmodes are to be
  determined and decompose the spectrum of $Q$ accordingly into desired
  eigenmodes, \beq {\cal W} =\mbox{spec}(Q)_w\; ,
\label{eq:window}
\eeq   
and the remainder:
\begin{equation}
\mbox{spec} (Q) = \mbox{spec}(Q)_w \cup \mbox{spec}(Q)_u\; .
\end{equation}
\item[-] Construct a polynomial $q$ that casts the spectral window
  $\mbox{spec}(Q)_w$ outside the region $[-1,1]$:
\begin{eqnarray}
\{ q( \lambda ) \vert \lambda \in \mbox{spec} (Q)_w \}
& \subseteq & [ -\infty , -1 [ \hspace{0.1cm} \cup 
\hspace{0.1cm} ] 1 , \infty ]\; , \\ \{ q( \lambda ) \vert \lambda \in
\mbox{spec} (Q)_u \} & \subseteq & [ -1, 1 ]\; .
\end{eqnarray}
\end{itemize}
We found that the simple choice
\begin{equation}
q(Q) = \frac{2}{s^2} Q^2 - (1+r)I
\label{eq:prep_pol}
\end{equation}
will accomplish the task, since the smallest eigenmodes of $q(Q)$, i.e.\ the
ones closest to $\lambda_{min}(q(Q))$, correspond to the
low-lying eigenmodes of $Q$. The polynomial carries two parameters:\\
(i) The scale factor $s$ is the spectral radius of $Q$.  It can be computed
in a first Arnoldi run on just a few vacuum configurations, since it
fluctuates little with the gauge field $\{U\}$. \\ (ii) The offset parameter
$r$ represents a simple shift operation and controls the actual size of the
window of desired eigenmodes.

\paragraph{Second step: polynomial acceleration}
The Arnoldi method is sensitive to the level density of eigenvalues,
$\rho(\lambda)$. For acceleration it is therefore crucial to
precondition the problem by decreasing the level density in the
spectral region of interest.  This can readily be achieved with
Chebyshev polynomials $T_N$ of degree $N$ due to their rapid increase
outside the interval [-1,1] (within which they are close to zero).\\

The practical procedure is then to find an appropriate Chebyshev polynomial
$T_N$ and compute the eigenmodes of largest moduli of 
$T_N\circ q\circ Q$, with $q$
taken from \eq{eq:prep_pol}.

We found the optimal convergence {\it w.r.t.}\ CPU time with a
Chebyshev polynomial $T_N$ of order $N= 80$ and $r$ chosen such that
approximately {\it only} the searched for 300 lowest-lying eigenvalues
were contained in the interval $[-1-r,-1]$. With this parameter
setting we needed just one Arnoldi
factorization~\cite{Golub:1996bk}. Since we chose the size of the
factorization to be 600, the calculation of the eigenmodes takes $2
\times 80 \times 600 = 96000 $ matrix-vector multiplications (where
the factor 2 is due to the fact that $Q^2$ enters the Chebyshev
polynomial).

We emphasize that without the Chebyshev acceleration step, we 
observed only very poor convergence.

Once we have the eigenfunctions, $| \psi_i \rangle$, of the operator
$T_N\circ q\circ Q$, it is straightforward to retrieve the eigenvalues
$\lambda_i$ of the original operator $Q$. By appealing to the relation
\beq Q| \psi_i \rangle = \lambda_i | \psi_i \rangle \Leftrightarrow
T_N(q(Q))| \psi_i \rangle =T_N(q(\lambda_i))| \psi_i \rangle \; , \eeq
we determine $\lambda_i$ through the Rayleigh quotient
\begin{equation}
\lambda_i =  \frac{\langle \psi_i | Q |
    \psi_i\rangle }{\langle \psi_i |
    \psi_i\rangle}\; .
\end{equation}
\begin{figure}[t]
\centering{\epsfig{figure=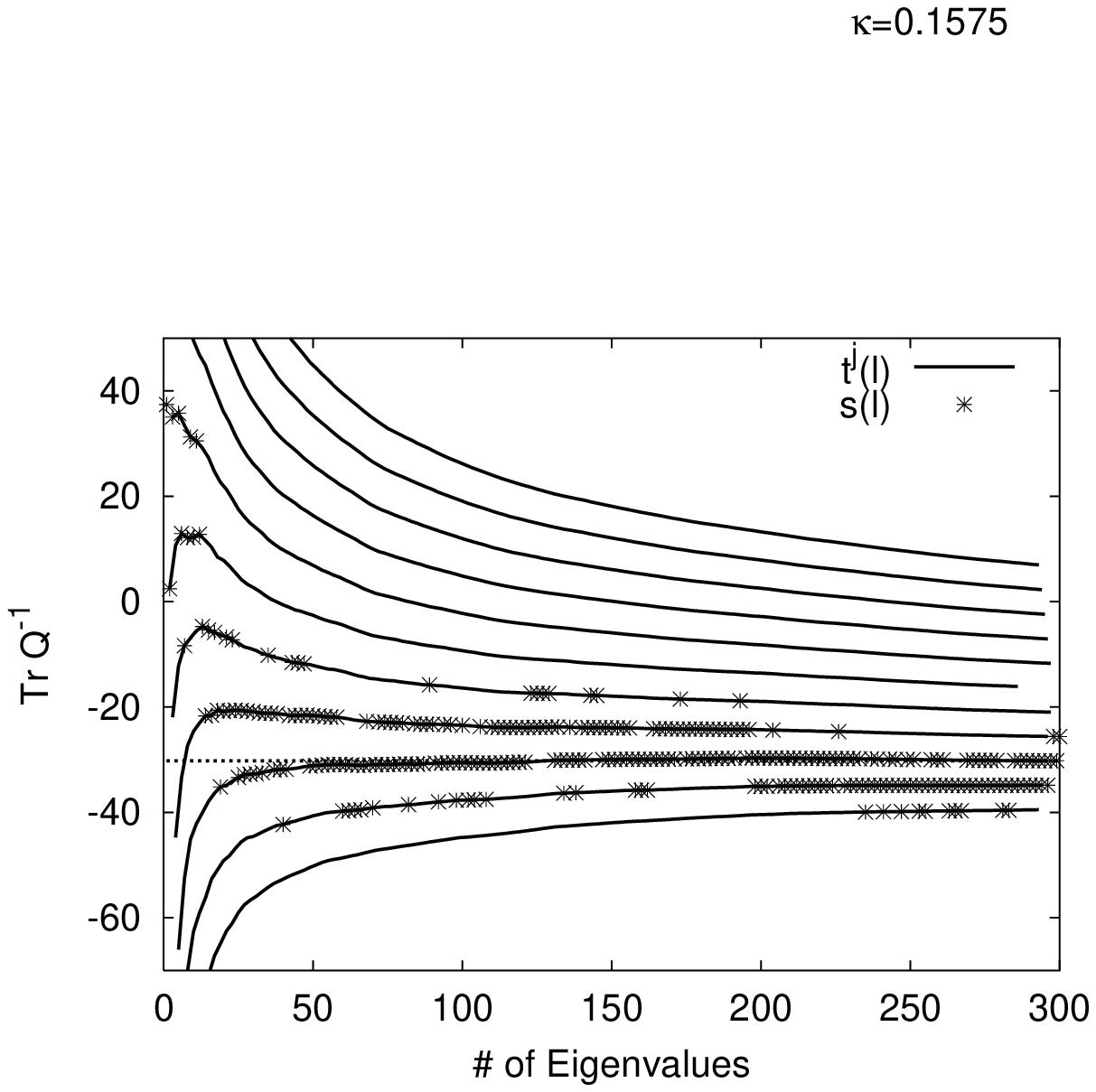,width=1.0\linewidth}} 
\vskip .5cm
\caption{Partial sum, showing for each $s(l)$ the associated  $j$ such that
 $s(l)=t^j(l)$.}
\label{fig:top5}
\end{figure}
\begin{figure}[t]
\centering{\epsfig{figure=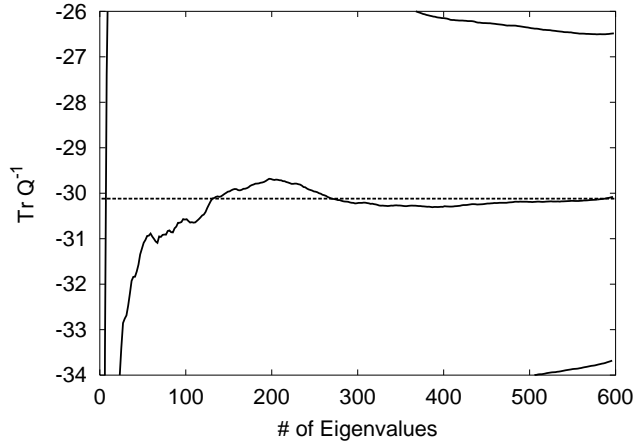,width=1.0\linewidth}} 
\vskip .5cm
\caption{High resolution plot of $t^p$ with 600 Eigenvalues}
\label{fig:top6}
\end{figure}

\begin{figure*}[ht]
\centering{\epsfig{figure=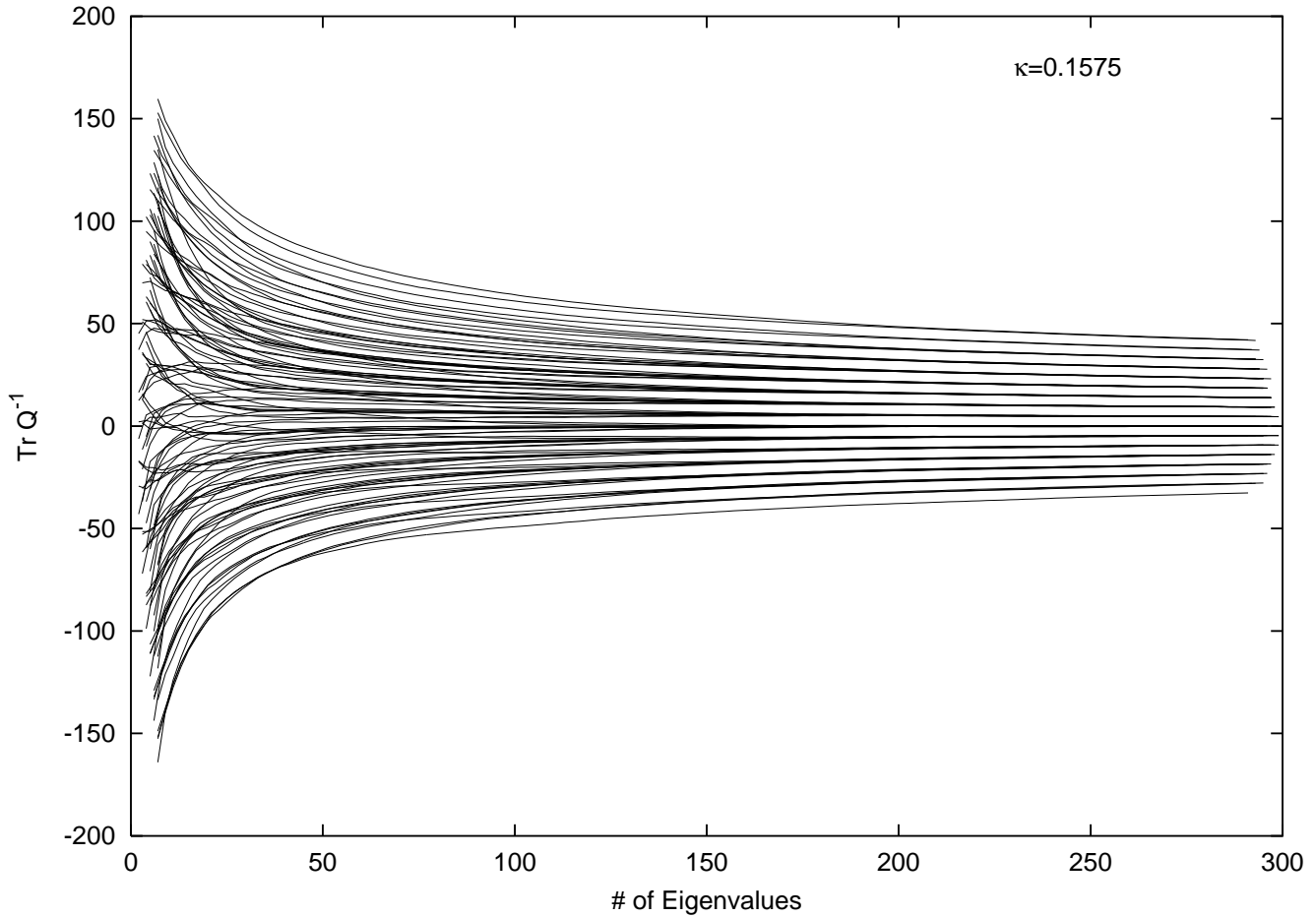,width=1.0\linewidth}} 
\vskip .5cm
\caption{Functions as in Fig.~\ref{fig:top2} for 10 gauge fields 
such that their $t$-values are all equal to 0. }
\label{fig:top3}
\end{figure*}
\begin{figure*}[ht]
\centering{\epsfig{figure=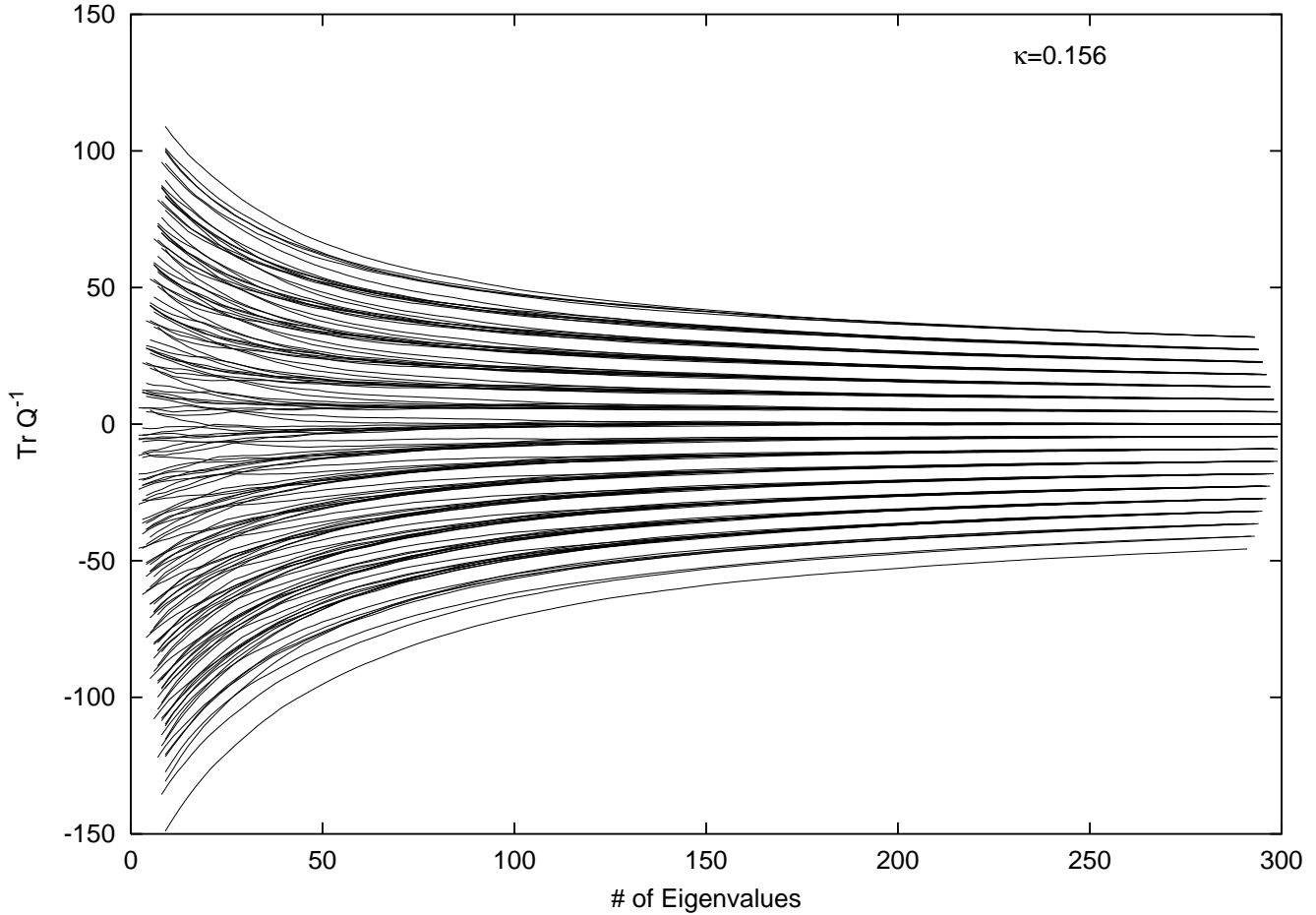,width=1.0\linewidth}} 
\vskip .5cm
\caption{Functions as in Fig.~\ref{fig:top3}, but for a heavier 
quark mass. }
\label{fig:top3_b}
\end{figure*}

\begin{figure}[t]
\centering{\epsfig{figure=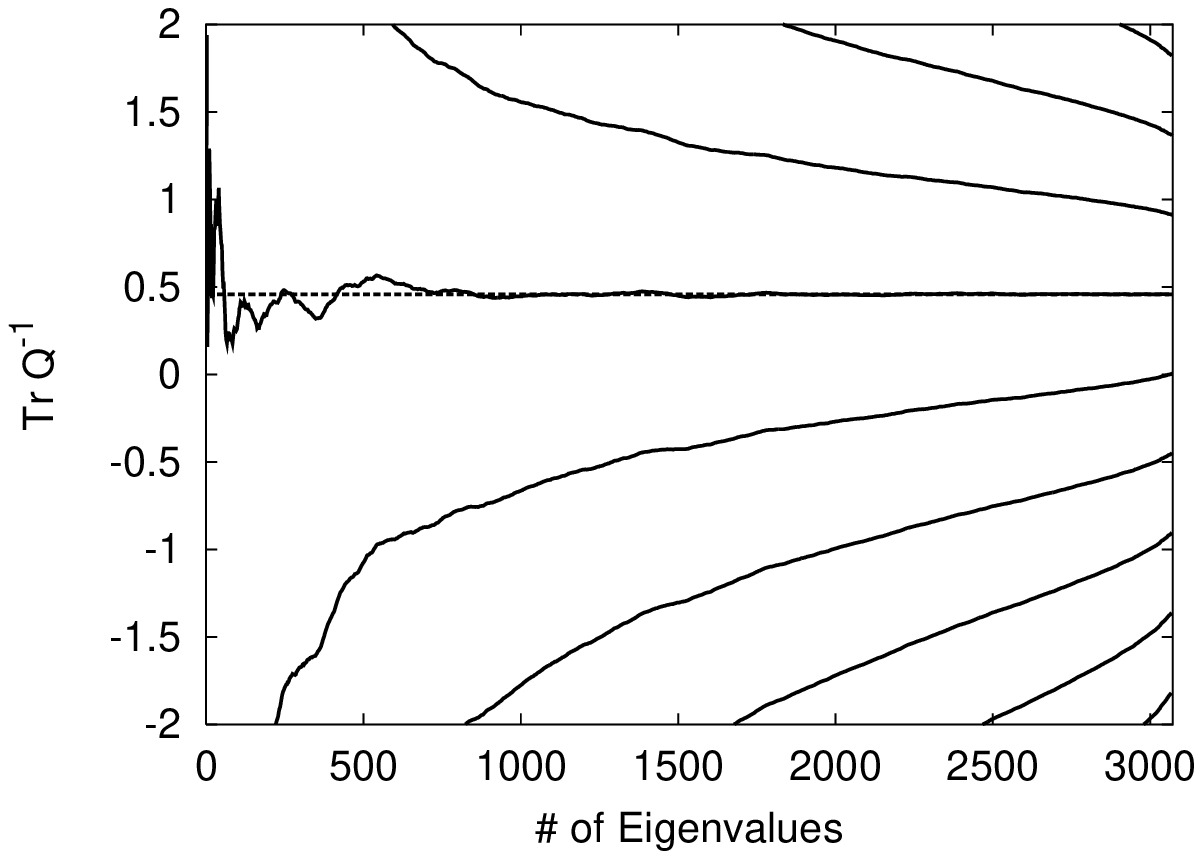,width=1.0\linewidth}} 
\vskip .5cm
\caption{Pattern of partial series $t^j$ on a quenched $4^4$ lattice at
  $\beta=5.0$, which allows for a calculation of the entire spectrum. }
\label{fig:top4}
\centering{\epsfig{figure=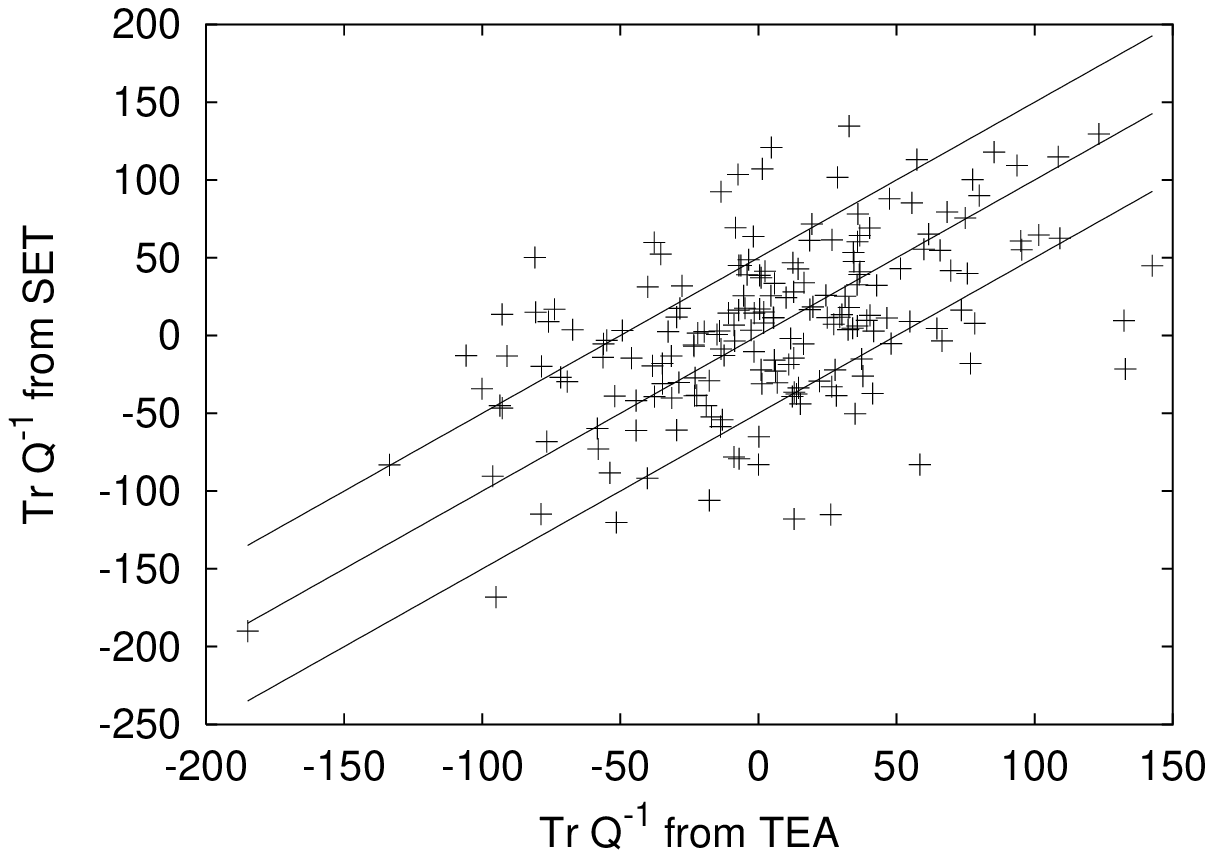,width=1.0\linewidth}} 
\vskip .5cm
\caption{Scatterplot comparing the results from TEA and SET
 on all configurations at $\kappa = 0.1575$.  The
errorband is chosen such that $67\%$ of the data points lie within.}
\label{fig:comp_tea_stoch}
\end{figure}
\begin{figure}[ht]
\centering{\epsfig{figure=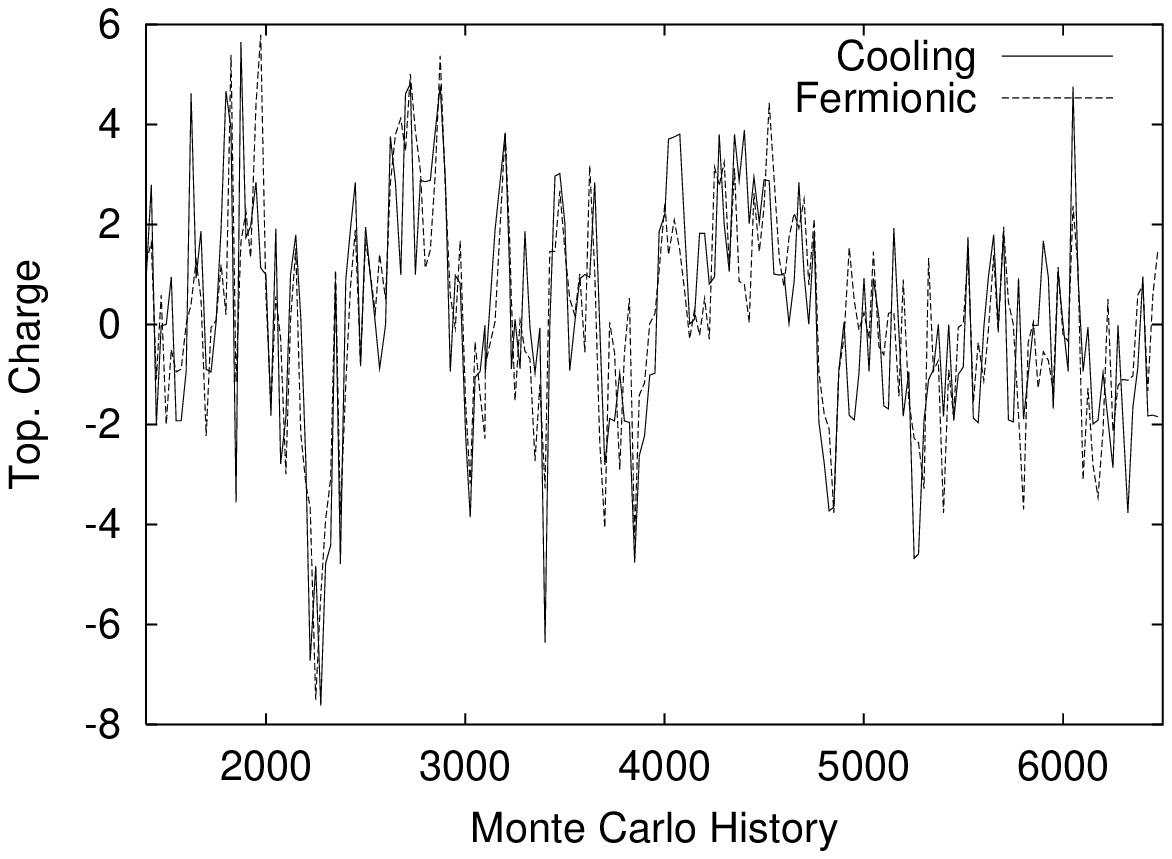,width=1.0\linewidth}} 
\vskip .5cm
\caption{Comparison of TEA's estimates of the topological charge with the
gluonic determination after cooling, on all SESAM configurations at lightest
quark
mass $\kappa = 0.1575$.}
\label{fig:top_cooling}
\end{figure}

Let us comment that -- for our present purposes -- we found the above
procedure to be much superior to the shift and invert
strategy~\cite{Golub:1996bk} which is frequently used to find the
low-lying eigenvalues

In \fig{fig:massdep} we illustrate the spectrum of $Q$ averaged over
the SESAM configurations, at lightest and heaviest sea quark mass
respectively, for which we plot the dependence of $|\lambda_i |$ on
$i$. The ordering is chosen according to \beq | \lambda_1 | \le |
\lambda_2 | \le | \lambda_3 | \le \ldots \quad .
\label{eq:nat_order}
\eeq The upper (lower) line corresponds to the heaviest (lightest) sea
quark mass of the SESAM sample. Their ratio, $| \lambda_i
(\mbox{lightest}) / \lambda_i (\mbox{heaviest})|$ is plotted in
\fig{fig:ratio} as a function of $i$. It shows a rather steep rise for
$i\le100$, which illustrates the growing importance of the
lowest-lying modes in the spectral representation of $Q^{-1}$,
\eq{eq:inverse}, when decreasing the sea quark mass.  This feature
should become even more pronounced for the imminent QCD simulations in
the yet deeper chiral regime, $m_{PS}/m_{V}< 0.5$.

\section{Determination of $\trq ^{-1}$ }\label{sec:bechmarking}

In the continuum limit, the complete trace 
\beq
{\cal Q} =\trq ^{-1}
\label{eq:trace}
\eeq is related via the index theorem~\cite{Atiyah:1971ws} to the net
topological charge of the gauge field.  Albeit the breaking of chiral symmetry
by the standard Wilson lattice action blurs this connection somewhat at finite
lattice spacings, $a$, one would still expect remnants of the index theorem to
hold~\cite{Smit:1987fn}.  In previous works, $\cal Q$ has been estimated by use
of stochastic estimator techniques (SET) which basically compute quark
propagators on stochastic sources, and indeed, a close
correlation between gauge field topology and $\trq ^{-1}$ has been
observed~\cite{Alles:1998jq,Bali:2001gk}.  Hence, it is a quantity of
significant interest.  This will be all the more the case in future
applications of the overlap fermion lattice
formulation~\cite{Neuberger:1999zk}.

$\cal Q$ is given by  the sum of the inverse eigenvalues:
\begin{eqnarray}
\trq ^{-1}&=& \mbox{Tr} \sum_i 
\frac{1}{\lambda_i} \frac{ | \psi_i 
\rangle \langle  \psi_i |} 
{\langle \psi_i |  \psi_i \rangle } \nonumber \\
&=& \sum_i \frac{1}{\lambda_i} \frac{\langle \psi_i |  \psi_i \rangle }
{\langle \psi_i |  \psi_i \rangle }= \sum_i \frac{1}{\lambda_i}\; . 
\label{eq:tcc}
\end{eqnarray}

Physically, it is the low-lying modes that encapsulate the interesting
information within the full sum of \eq{eq:tcc} while the large modes provide
just nuisance by adding background noise to the infrared signal.  {\it Heu me
  miserum}, in order to turn this qualitative proposition into a quantitative
statement we would need to know the transition point between infrared and
ultraviolet physics. For an approximate use of the spectral relation,
\eq{eq:tcc}, we need a cunning technique to deplete the unwieldy background.

With this in mind let us take a heuristic approach and study the
pattern of eigenvalue distributions as obtained from the SESAM 
ensemble of vacuum configurations. To reach
our goal, it is useful to order the spectrum and perform certain
partial summations in \eq{eq:tcc}:

Let $p_i$ denote the positive and $n_i$ the negative eigenvalues,
\begin{equation}
\{ \lambda_i \} = \{ p_i \} \cup \{ n_i \},
\end{equation}
and let them be ordered such that:
\begin{eqnarray}
& \hspace{.11cm}p_1 \le p_2 \le p_3 \le \ldots & \\
& | n_1 | \le | n_2 | \le | n_3 | \le \ldots &
\quad . \label{eq:order}
\end{eqnarray}

In order to study convergence properties let us define the following partial
series:
\begin{eqnarray}
t^j(l)
&=&\sum_{i=1}^{j+k} \frac{1}{p_i}+\sum_{i=1}^{k} \frac{1}{n_i},
\hspace{.5cm} j \ge 0 \label{eq:series1} \\
t^j(l)
&=&\sum_{i=1}^{j+k} \frac{1}{n_i}+\sum_{i=1}^{k} \frac{1}{p_i},
\hspace{.5cm} j < 0 \label{eq:series2}\\
l &=& |j| + 2k\; ,
\end{eqnarray}
where, in obvious notation, the parameter $j$ labels the excess number of
entries with positive (negative) over the ones with negative (positive)
eigenvalues.  In \fig{fig:top1} and \fig{fig:top2} the values of $t^j$ are
plotted for the index range $-5 \le j \le 5$, as obtained from {\it one
  particular} SESAM configuration at our lightest available quark mass,
$\kappa = 0.1575$.  The partial sums appear to exhibit a certain convergence
pattern which is displayed in \fig{fig:top2}, where we connected the points to
given $j$-values.

Alternatively, one might organize the summation in the order of
increasing moduli of the eigenvalues (\eq{eq:nat_order}), independent
of their signs, and define the truncated sums \beq s(l)=\sum_{i=1}^{l}
\frac{1}{\lambda_i}\; .
\label{eq:nat_order_sum}
\eeq The family of curves, ${\cal F} = \{t^j(l)\}$ provides a suitable
framework to disclose the asymptotic behavior
of this
inverse eigenvalue summation, $s(l)$.  This is illustrated, again for the
particular gauge configuration, in \fig{fig:top5}, where we display the data
points for $s(l)$ in the range $1 \leq l \leq 300$. For reference we also show
the (slowly narrowing) band of the partial sums, $t^j(l)$.  Note that
from $s \approx 150$ onward, $s(l)$ in this particular configuration
jumps\footnote{ To put it differently: the asymptotic distribution of
  eigenvalues is characterized by alternating signs when proceeding according
  to the order given by \eq{eq:nat_order}.}  mostly between the two levels
$t^{-3}$ and $t^{-4}$.  Moreover $t^{-3}$ appears to be distinguished as
lying between the asymptotically falling set of curves with $j \geq -2$
and the rising ones, $j \leq -4$.

We found this scenario to apply {\it mutatis mutandis} to all configurations
in the sense that
\begin{itemize}
\item for each gauge field $\{U\}$, $\cal F$ contracts around a
particular partial sum $t^p(l)$ that levels to a plateau value beyond
$l\approx 150$, with $p$ depending on the choice of the gauge field
$[U]$.
\end{itemize}

Let us next quantify our observations and denote the height of this supposed
plateau with $t$, setting 
\beq
t := t^p(300)\; .
\label{eq:plateau}
\eeq

Note that $t$ varies with the underlying gauge configuration
$[U]$. The question then arises how accurately we can extract the
actual plateau height.  In order to appreciate the numerical flatness
of the plateau curve we pushed the eigenvalue computation to a number
of 600 modes {\it on a single} SESAM configuration.  In
\fig{fig:top6} we display the resulting plateau on a magnified
scale. Assuming that the apparent remaining weak oscillation for
$l>150$ is a {\it pars pro toto} feature for the entire SESAM sample
and that with 300 eigenvalues one has already passed the first
extremum on the entire sample of gauge fields $U$, we estimate that
from the lowest-lying 300 eigenvalues $t$ can be determined with a
{\it bona fide} accuracy of $1\%$.

\begin{figure}[t]
\centering{\epsfig{figure=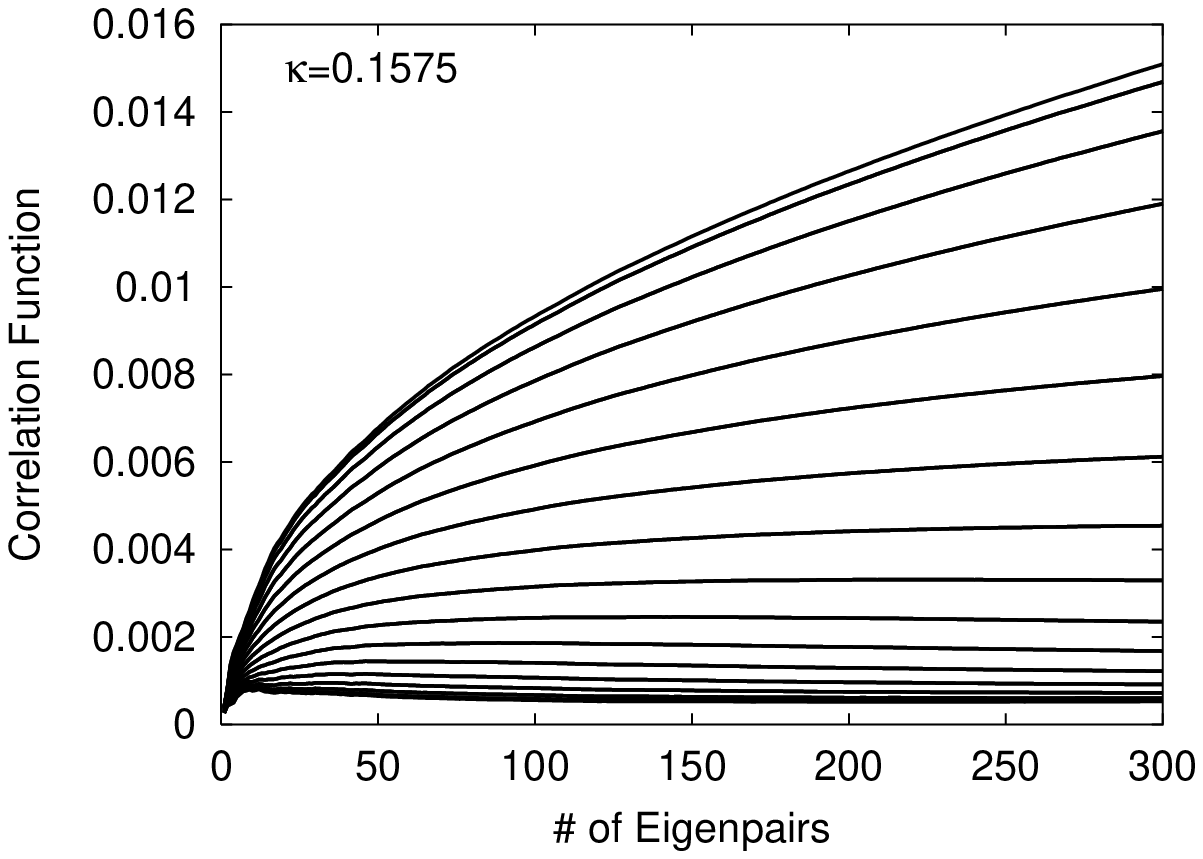,width=1.0\linewidth}} 
\vskip .5cm
\caption{Set of $\pi$  correlation function $C^l(\Delta t)$ from TEA on
  local sinks and sources, plotted against the spectral cutoff, $l$.  $\Delta
  t $ increases as one steps down from the top curve which refers to $\Delta
  t = 0$.  }
\label{fig:corr1}
\centering{\epsfig{figure=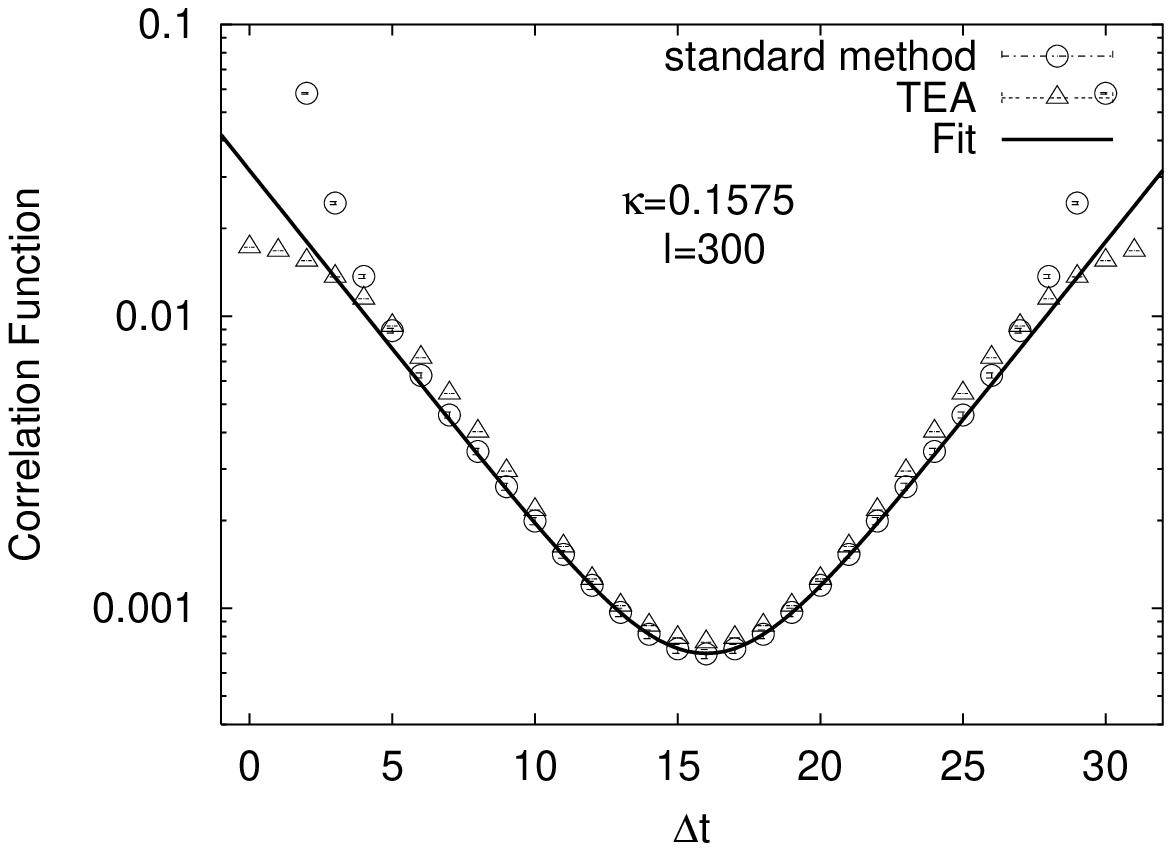,width=1.0\linewidth}} 
\vskip .5cm
\caption{Comparison of the $\pi$ correlation function as provided by
  TEA with $l = 300$, with the one obtained from the standard method (solving
  linear systems) on local sinks and sources.}
\label{fig:corr2}
\end{figure}
\begin{figure}[t]
\centering{\epsfig{figure=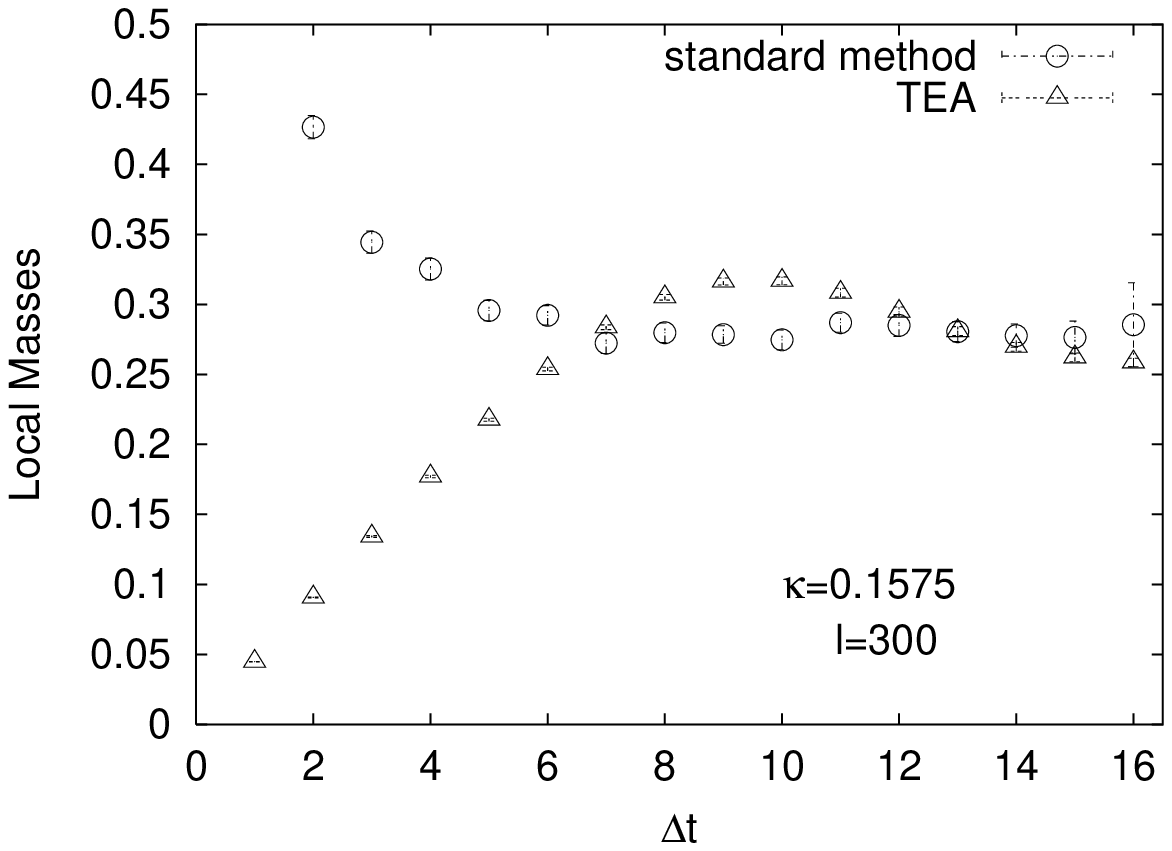,width=1.0\linewidth}} 
\vskip .5cm
\caption{Local $\pi$ masses from TEA and the standard method .}
\label{fig:corr3}
\centering{\epsfig{figure=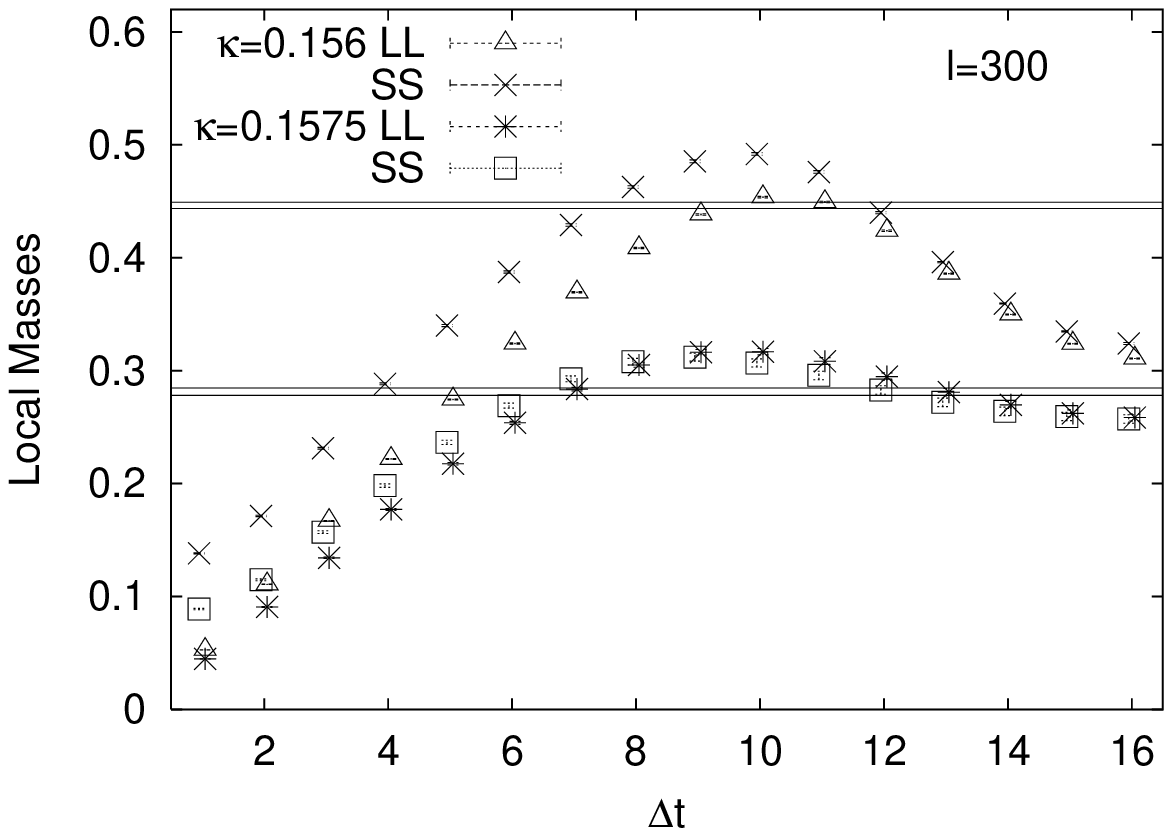,width=1.0\linewidth}}
\vskip .5cm
\caption{Local masses of the $\pi$ from TEA.  LL
  are the unsmeared results, whereas SS
  stands for smearing the sinks and sources.
  The horizontal lines show the errorbands of the $\pi$ as
  obtained from the standard method (solving linear systems).}
\label{fig:corr4}
\end{figure}
\begin{figure}[t]
\centering{\epsfig{figure=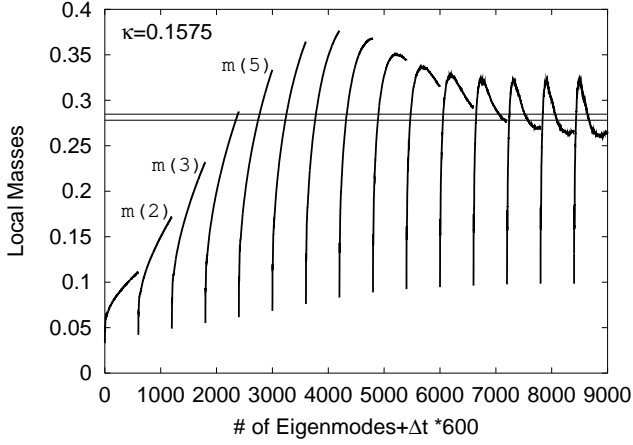,width=1.0\linewidth}} 
\vskip .5cm
\caption{The dependence of the local masses $m(\Delta t)$ on the
  spectral cutoff, $l \leq l_{max}= 600$, on a single configuration. The
  x-axis carries the parameter $l_{\Delta t}=600 \Delta t +l$ with $\Delta t =
  1,2, \ldots , 15$.  The horizontal line shows the errorband of the $\pi$
  mass as obtained from the standard method (solving linear systems).}
\label{fig:corr5}
\end{figure}
\begin{figure}[t]
\centering{\epsfig{figure=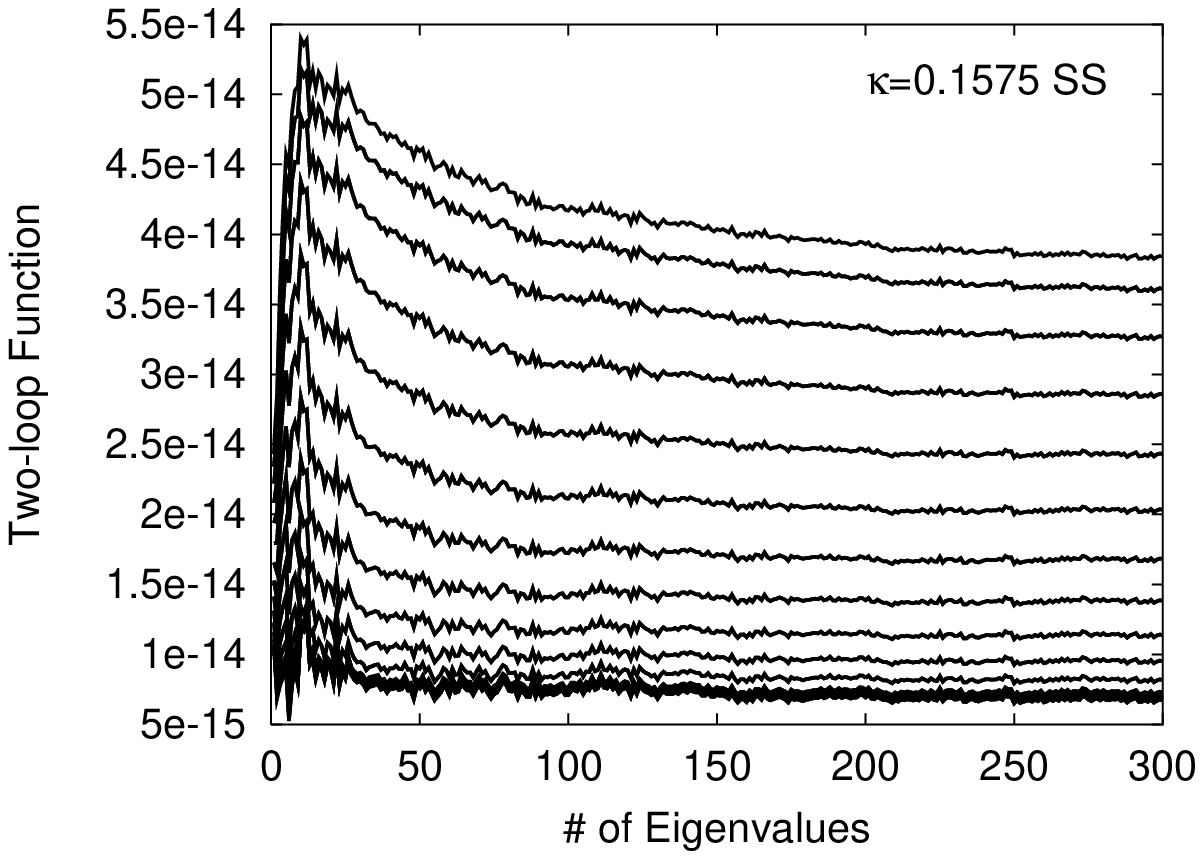,width=1.0\linewidth}} 
\vskip .5cm
\caption{Set of two-loop functions $T^l(\Delta t)$ on smeared sources
and sinks plotted versus the cutoff $l$.  As in Fig.~\ref{fig:corr1},
$\Delta t$ increases as one steps down from the top curve.}
\label{fig:tl1}
\end{figure}

Next let us argue that the plateau value, $t[U]$, provides us with an
approximant for the complete trace, \eq{eq:trace}: 
\beq {\cal Q}[U] \approx t[U] \; .
\label{eq:tr_approx}
\eeq What is the deviation from the complete trace in our situation?
In our range of quark masses, we can exclude zero-level crossings of
eigenmodes. Therefore the matrix $Q$ possesses an equal number of
positive and negative eigenvalues. Hence, when adding up all $n$
eigenvalues (with $n = \mbox{dim}(Q)$), $s(n)$ will lie on the curve
with superscript 0
\begin{equation}
{\cal Q} = s(n) = t^0 \left(n \right), 
\hspace{.2cm} \mbox{with } n= \mbox{dim}(Q)\; .
\end{equation}
Thus $\trq ^{-1}$ is related to the plateau height 
$t$  in the following way:
\begin{equation}
{\cal Q}  = t + \frac{p}{\left|\lambda_n \right|}\; .
\label{eq:t_asympt}
\end{equation}
The second term on the {\it r.h.s.}\ measures the distance between
$t^0$ and $t^p$ at $l = n$. It can be neglected with respect
to the error on $t$ itself, since $|1 / \lambda_n \approx .15|$ is
approximately equal to the error of $t$.

Another justification for the validity of our approximation,
\eq{eq:tr_approx}, comes from our observation that different field
configurations $[U_r]$ yield equal results when plotted with appropriate
offsets $t[U_r]$, namely $(t^j(l)[U_r]-t[U_r]) $.  This is illustrated in
\fig{fig:top3} and \fig{fig:top3_b} 
where we superimpose 10 such series for $U_r (r=1,\ldots 10)$:
it strikes the eye that, for $l>150$, the partial sums to the 10 gauge fields
all {\it collapse onto a single, universal} family of curves:
\begin{equation}
t^{j+p_r}(l)[U_r]-t[U_r]=t^{j+p_s}(l)[U_s]-t[U_s]\; .
\end{equation}
Hence there exists a set of  $U$-independent functions $\hat{t}^j(l)$ such that
the following identity with respect to $U$ applies:
\begin{equation}
t^{j+p[U]}(l)[U]-t[U]=\hat{t}^{j}(l)\; .
\end{equation}

This pattern strongly supports the picture that on our configurations the
interesting physics with respect to the topological charge is indeed contained in
the subset of the 150 smallest eigenvalues, while the remaining ones carry
no information on $\cal Q$.

We corroborate this result by considering a $4^4$ lattice where we
determined all 3072 eigenvalues of $Q$ in quenched QCD at $\beta =
5.0$.  In \fig{fig:top4} we plotted the corresponding partial sums
$t^j$.  Notice that $t^p$ (here $p$ happens to be 0) remains
absolutely flat after reaching its plateau value at around $l \simeq
500$.

A comparison of the $\cal Q$ values as produced by TEA with the ones
obtained in Ref.\ \cite{Bali:2001gk} from SET on the entire $\kappa =
0.1575$ sample is shown in \fig{fig:comp_tea_stoch}. The data points
seem to scatter rather nicely around the bisecting line. While the
accuracy of the TEA results on individual configurations is about
$1\%$, the uncertainty of the SET estimates turns out to be $\Delta
{\cal Q} \approx \pm 50$.

In \fig{fig:top_cooling} we present TEA's (normalized) $\cal Q$ values
along the Monte Carlo history of our SESAM sample at $\kappa = 0.1575$
and compare them to the result of the gluonic determination after
cooling~\cite{Bali:2001gk}.  We reconfirm our previous
finding~\cite{Alles:1998jq} that there is a close correlation between
the gluonic and fermionic definitions of the net topological charge.

\section{Hadronic two-point functions}

In this section we wish to investigate the potential of spectral
methods in the computation of two-point hadronic
correlators~\cite{DeGrand:2000gq}.  The question here is to what
extent we can verify low-lying eigenmode dominance for infrared
physics, specifically the ground states in particular hadronic
channels.

In order to set the stage we shall first consider the octet pseudoscalar
channel as we can easily compare to standard pion correlator computations.
We shall then elaborate on the spectral approach to the singlet
pseudoscalar propagator which differs from the octet one by the
two-loop correlator. The latter involves the computation of incomplete
trace expressions $\langle {\cal Q}(t){\cal Q}(t+\Delta t) \rangle$ and
thus represents a quantity of increased complexity.

\subsection{Basics for mass determinations}

Masses are extracted  from the large time behavior of
correlation functions $C(t)$ that carry the quantum numbers of the 
particles in question. The correlation functions of the $\pi$ and
the $\eta'$ have the following form \newpage
\begin{eqnarray}
C_{\pi} (\Delta t) &=&\Big< \sum_{s_i,\alpha_i,a_i,t} \big[ Q^{-1}
(s_1,t,\alpha_1,a_1;s_2,t+\Delta t,\alpha_2,a_2) \nonumber \\
&&Q^{-1}(s_2,t+\Delta t,\alpha_2,a_2;s_1,t,\alpha_1,a_1) \big] \Big>_U
\label{eq:pc} \\ C_{\eta'} (\Delta t) &=& C_{\pi} (\Delta t) - N_f
\label{eq:eceq} \Big< \sum_t {\cal Q}(t){\cal Q}(t+\Delta t)
\Big>_U, \nonumber
\end{eqnarray}
with ${\cal Q}(t)= \sum_{s,\alpha,a}
Q^{-1}(s,t,\alpha,a;s,t,\alpha,a)$ and $N_f=2$ flavors.  The
coordinate $n$ is subdivided into a spatial $s$ and a temporal part
$t$.  By introducing energy eigenfunctions one can show that these
correlation functions decay exponentially in time, with the particle
mass being the decay constant $C(\Delta t) \sim \mbox{exp}(-m \Delta
t)$. On a toroidal lattice however with temporal extent $T$ this
exponential decay appears as a cosh, $C(\Delta t) \sim
\mbox{exp}(-m\Delta t)+\mbox{exp}(m(\Delta t-T))$.  Local masses $m$
can be retrieved for every value of $\Delta t$ by solving the implicit
equations
\begin{equation}
\frac{C(\Delta t+1)}{C(\Delta t)}= 
\frac{\mbox{exp}(-m(\Delta t+1))+\mbox{exp}(m(\Delta t+1-T))}
{\mbox{exp}(-m\Delta t)+\mbox{exp}(m(\Delta t-T))}\; ,
 \label{eq:localmasses}
\end{equation}
with respect to $m$. 
Plateaus in the time dependence of the local masses $m(\Delta t)$
exhibit the masses of the particles.

\subsection{ Smearing}
The extension of the local mass plateaus can be increased by
enhancing the overlap with the  ground state 
correlation function. We follow Ref.~\cite{Gusken:1990qx} 
and apply the  smearing matrix $S=(F)^k$, with 
\begin{eqnarray}
F(n,\alpha,a;m,\beta,b)= \frac{1}{1+6l}
\Big[ \delta(n,\alpha,a;m,\beta,b) \\
 + l \sum_{\mu=1}^3 
\Big( 
U_{n,\mu}(a;b)\delta(n+\mu;m) \nonumber\\
 + U^{\dagger}_{n-\mu,\mu}(a;b)\delta(n-\mu;m)  
\Big) \Big] \; ,\nonumber
\end{eqnarray}
choosing   $k=50$ and $l=4$.

Source and sink smearing are readily  accomplished
by the replacements
\begin{equation}
 \psi_i \rightarrow \psi_i^s = S \psi_i
\end{equation}
in the spectral propagator representation
\begin{eqnarray}
&Q^{-1}_{sm}&(n,\alpha,a;m,\beta,b) = \nonumber\\
&&\sum_i\frac{1}{\lambda_i} \frac{ | \psi_i^s (n,\alpha,a)
\rangle \langle \psi_i^s (m,\beta,b) | } 
{ \langle \psi_i | \psi_i \rangle }\; .
\label{eq:smear_prop}
\end{eqnarray}

\begin{figure*}[ht]
\centering{\epsfig{figure=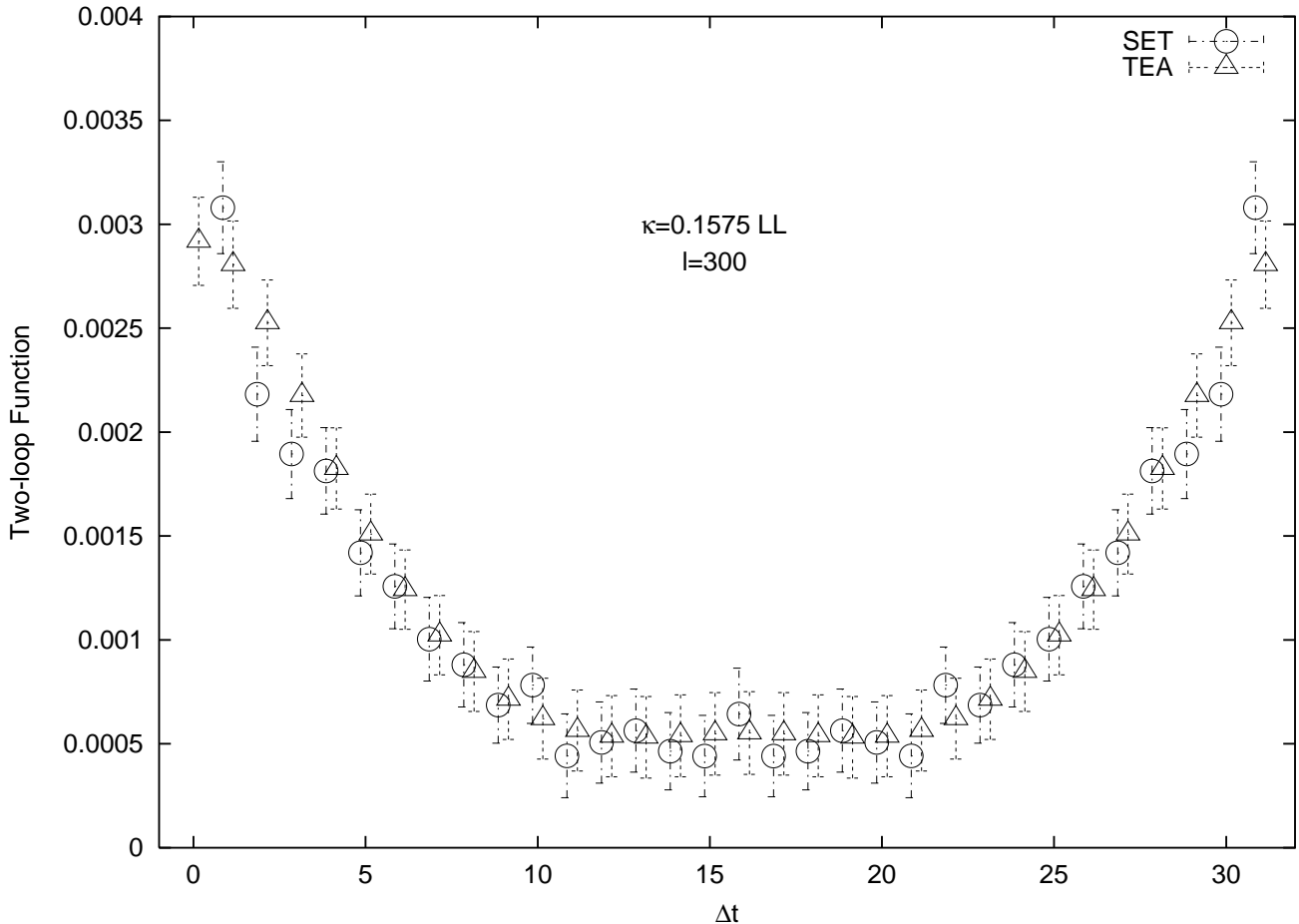,width=1.0\linewidth}} 
\vskip .5cm
\caption{Two-loop correlator (Eq.~\ref{eq:tlf}), estimated with TEA and
SET, for the lightest quark mass and local sources and sinks.}
\label{fig:tl2}
\end{figure*}

Throughout this section we will assume the eigenmodes to be
enumerated  according to \eq{eq:nat_order}.

\begin{figure*}[ht]
\centering{\epsfig{figure=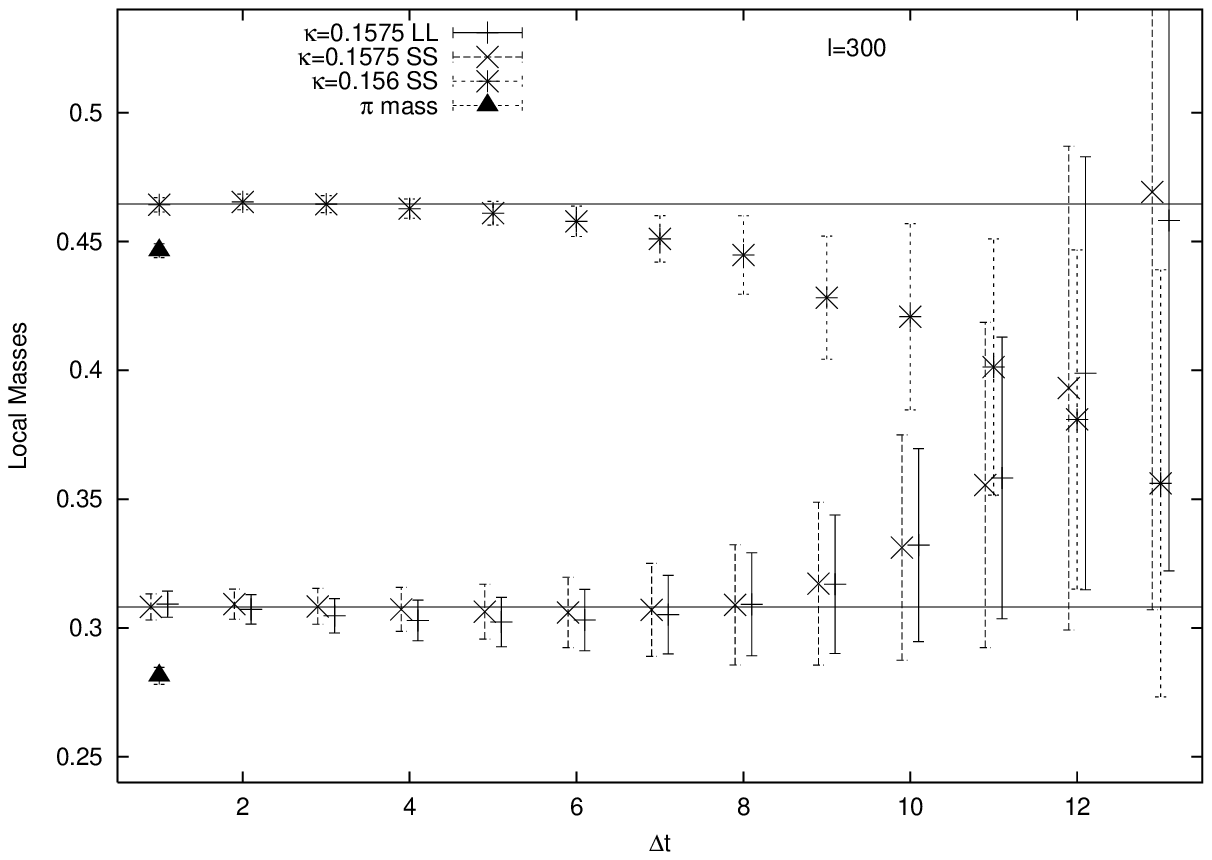,width=1.0\linewidth}} 
\vskip .5cm
\caption{Local $\eta'$ masses from TEA on local (LL) and smeared (SS) sources
  and sinks with ground state projection of its connected piece (OLGA). For
  comparison the $\pi$ mass as obtained from the standard method is plotted.}
\label{fig:tl3}
\end{figure*}

\subsection{Pion correlator}

The spectral representation of $C_{\pi}(\Delta t)$ reads
\begin{eqnarray}
C_{\pi}(\Delta t) &=& 
\sum_{i,j,t} \frac{1}{\lambda_i \lambda_j}
\frac{ \langle \psi_i(t) | \psi_j(t) \rangle 
\langle \psi_j(t+\Delta t) |  \psi_i(t+\Delta t) \rangle} 
{ \langle \psi_i | \psi_i \rangle 
\langle  \psi_j | \psi_j \rangle } \nonumber \\
&\equiv& \sum_{i,j} \Psi(i,j,\Delta t)\; ,
\label{eq:pico}
\end{eqnarray} 
where we suppressed the brackets that represents the
average over the gauge fields.  

For $C_{\pi}$ no early saturation over the entire $\Delta t$-range
can be expected, as one can see by integrating \eq{eq:pico}
\begin{equation}
\sum_{\Delta t} C_{\pi} (\Delta t) = \sum_i \frac{1}{\lambda_i^2}\; .
\label{eq:pcs}
\end{equation} 
Since all the contributions on the {\it r.h.s.}\ of \eq{eq:pcs} are
positive the series is monotonically increasing. Therefore -- contrary
to the case of the topological charge -- neither cancellation effects
nor early saturation can be expected in this global quantity.  But
what about the regime of infrared physics described by the correlator,
i.e.\ its asymptotic behavior in $t$?  

Let us consider the truncated
spectral correlator
\begin{equation}
C_{\pi}^{l}(\Delta t) =\sum_{i,j=1}^l \Psi(i,j,\Delta t)\; .
 \label{eq:hopeless}
\end{equation}
In order to demonstrate the low-lying eigenmode dominance at large time
separations we present in \fig{fig:corr1} a family of curves,
$C_{\pi}^{l}(\Delta t)$, for the various time slices $\Delta t$, plotted
against the spectral cutoff, $l$, at the lightest SESAM quark mass.  It is
gratifying to find that $C_{\pi}^l (\Delta t)$ for $\Delta t \ge 7$ shows a
flat behavior in the regime $l > 100$.  On the other hand for small time
separations higher eigenmodes continue to add -- in accordance with the idea of
excited state contaminations.

It is interesting to carry out a direct comparison of TEA with the standard
propagator as computed by linear solvers on a local source, in order to see
saturation occur in the region of interest, see \fig{fig:corr2}. We find good
agreement in the asymptotic regime, $7 \le \Delta t \le 25$.  We also show the
fit (fit range:$[8,15]$) to the data from the inverter
to the usual
$\cosh$ parameterization:
\begin{equation}
C^g_{\pi} (\Delta t) = A\; \mbox{cosh} \left[ m_{\pi} 
( \Delta t - T/2 ) \right]\; . \label{eq:fit}
\end{equation}

A much more sensitive test of TEA is to look at local effective masses.  In
\fig{fig:corr3} we compare, at the cutoff value $l = 300$, the TEA results with
the ones from standard propagator analysis, for the lightest SESAM quark mass.
The effects of smearing and varying quark masses are displayed in
\fig{fig:corr4}, again for $l= 300$.  As anticipated, we do observe a clear
tendency for improvement in the spectral approach with decreasing quark mass.
Yet there remains a marked oscillatory behavior over the entire SESAM range
of quark masses. 
Moreover we notice that smearing slightly improves the signal. 

A synopsis on the cutoff dependence of $m(\Delta t)$ is presented in
\fig{fig:corr5}, as obtained on a particular configuration at the lightest
quark mass.  To avoid cluttering of the data we have spread out the different
curves by means of the variable $l_{\Delta t} = 600 \Delta t + l$.  This
survey plot is meant to convey an idea how the oscillation will dampen out
with increasing cutoff $l$. 

Thus we have demonstrated that in the sea quark mass regime of the
SESAM configurations, there is insufficient dominance of the low-lying
eigenmodes to utilize TEA for a sensible calculation of the $\pi$ mass.

\subsection{$\eta'$-correlator}

Let us consider next the flavor singlet pseudoscalar channel with the
ground state particle $\eta'$.  As described in \eq{eq:pc} $C_{\eta'}$
differs from $C_{\pi}$ by the two-loop correlator $T$:
\begin{equation}
\label{eq:etas}
C_{\eta'}(\Delta t)=C_{\pi}(\Delta t)- N_f T(\Delta t)\; ,
\end{equation}
the spectral representation of which reads
\begin{eqnarray}\label{eq:tlf}
&&T(\Delta t) =  \\
&&\sum_t \sum_i \frac{1}{\lambda_i}
\frac{ \langle \psi_i(t) | \psi_i(t) \rangle}
{ \langle \psi_i | \psi_i \rangle} 
\sum_j \frac{1}{\lambda_j}
\frac{\langle \psi_j(t+\Delta t) | \psi_j(t+\Delta t) \rangle}
{\langle \psi_j | \psi_j \rangle}\; , \nonumber 
\end{eqnarray}
where we suppressed again the brackets indicating
the average over the gauge fields.

Again we sum $T(\Delta t)$ over $\Delta t$ in order to 
learn about TEA's potential in the two-loop situation 
\begin{equation}
\sum_{\Delta t} T(\Delta t)= \left( \sum_i \frac{1}{\lambda_i}
\right)^2 \label{eq:tl}\; .
\end{equation}
This is just the square of the 'topological charge', see  \eq{eq:tcc}.
Therefore we might  expect  TEA to 
work as well
as in section
\ref{sec:bechmarking}.

\begin{figure}[t]
\centering{\epsfig{figure=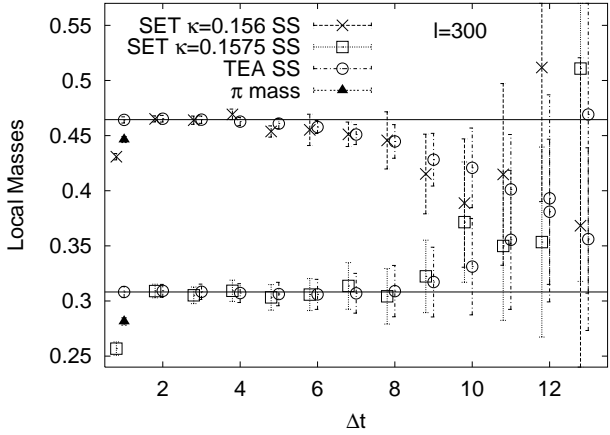,width=1.0\linewidth}} 
\vskip .5cm
\caption{Comparison of the local OLGA $\eta'$ masses from TEA and from
Stochastic Estimations on smeared sources and sinks.}
\label{fig:compstoch}
\end{figure}
\begin{figure}[t]
\centering{\epsfig{figure=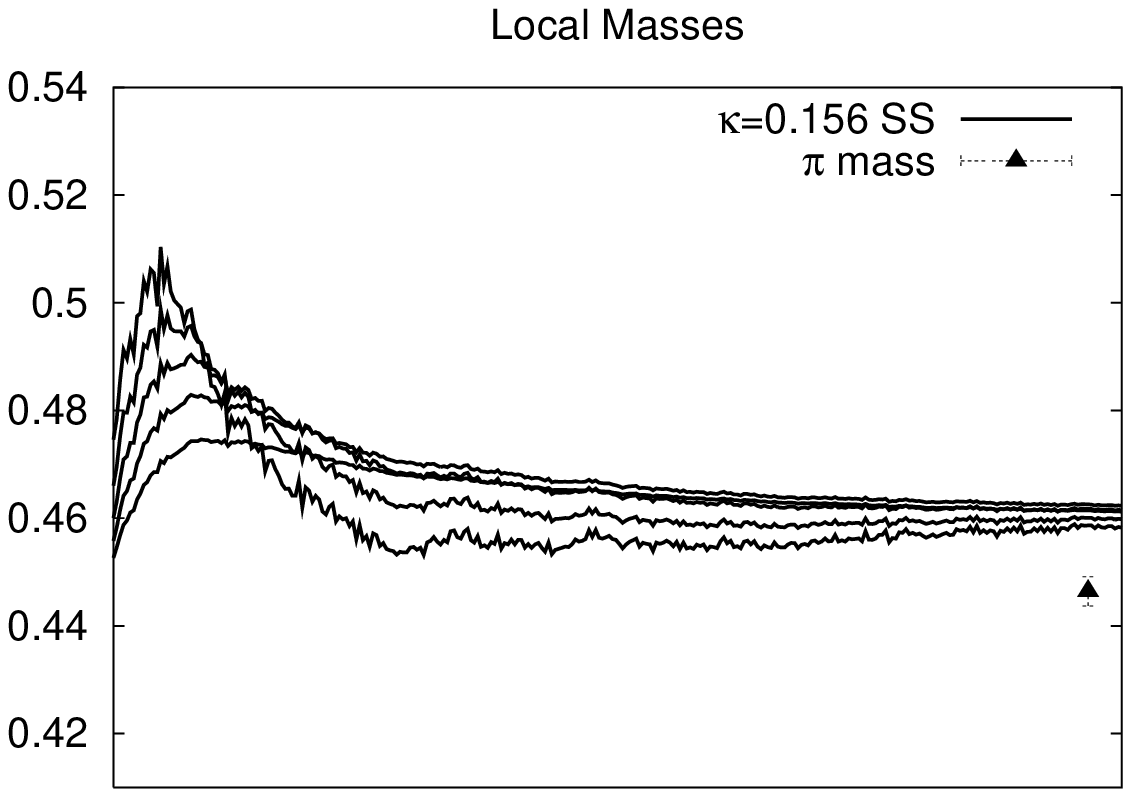,width=1.0\linewidth,height=0.6\linewidth}} 
\label{fig:tl4}
\vspace{-0.9cm}
\centering{\epsfig{figure=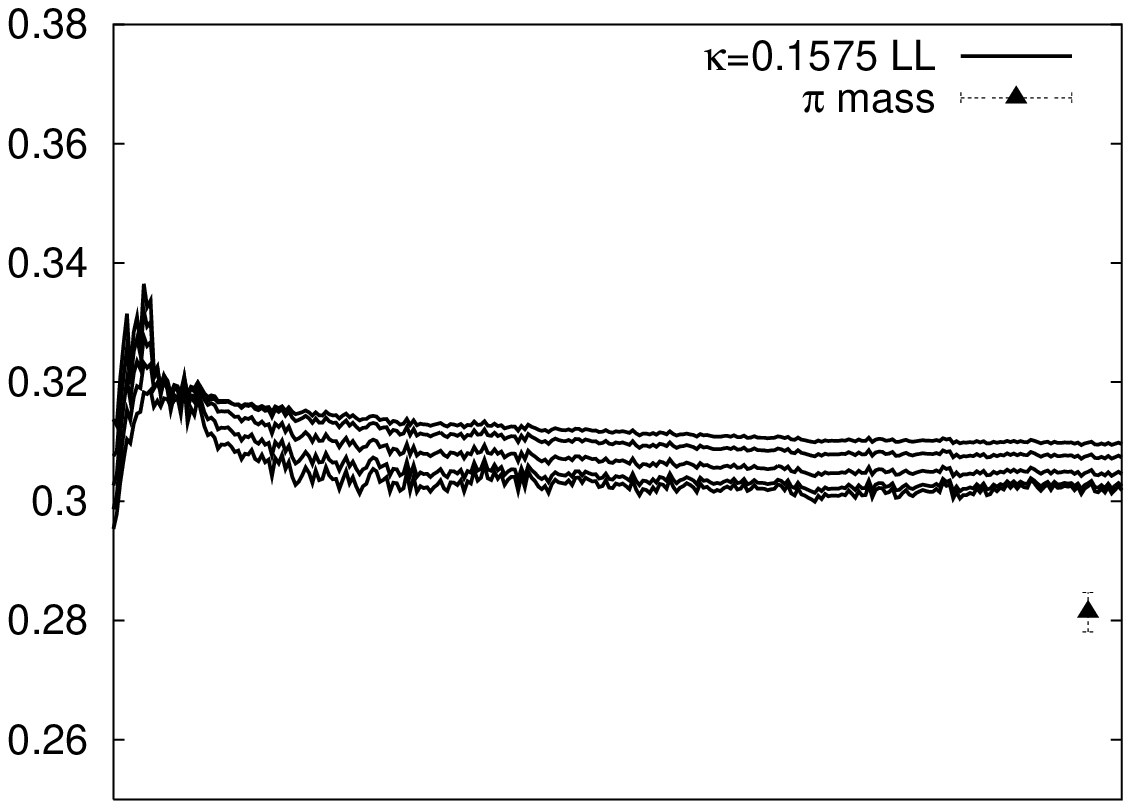,width=1.0\linewidth,height=0.6\linewidth}} 
\label{fig:tl5}
\vspace{-0.9cm}
\centering{\epsfig{figure=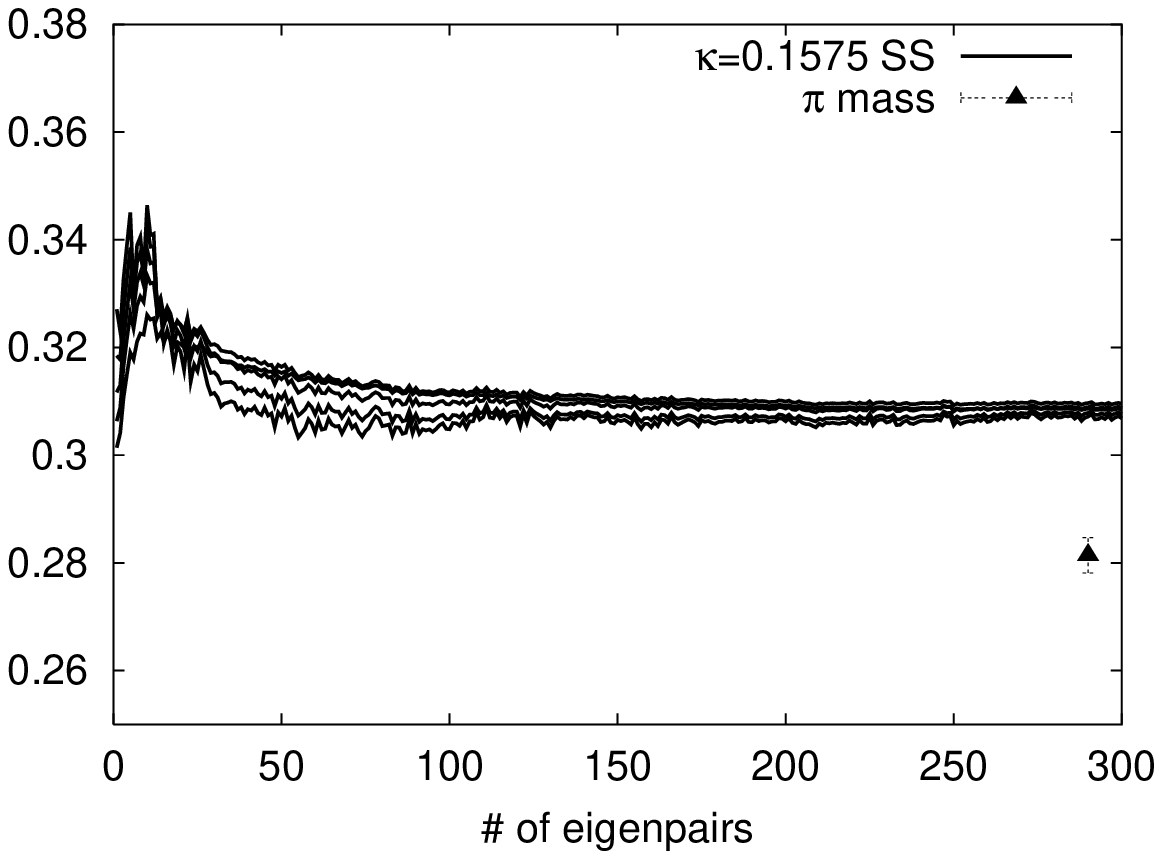,width=1.0\linewidth,height=0.6\linewidth}} 
\vskip .5cm
\caption{Dependence of the local $\eta'$ masses $m(\Delta t)$ for $\Delta
t=1,2,\ldots 5$ on the spectral cutoff.}
\label{fig:tl6}
\end{figure}
\begin{figure}[t]
\centering{\epsfig{figure=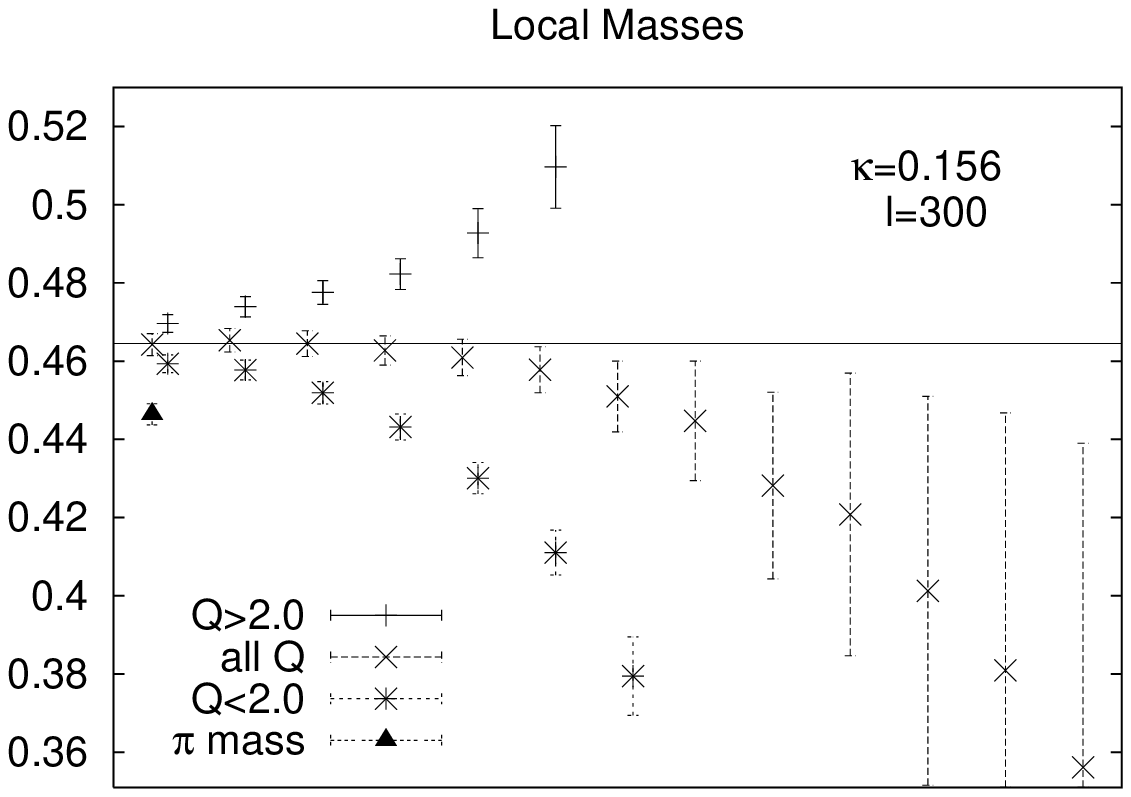,width=1.0\linewidth}} 
\label{fig:tl9}
\vspace{-0.95cm}
\centering{\epsfig{figure=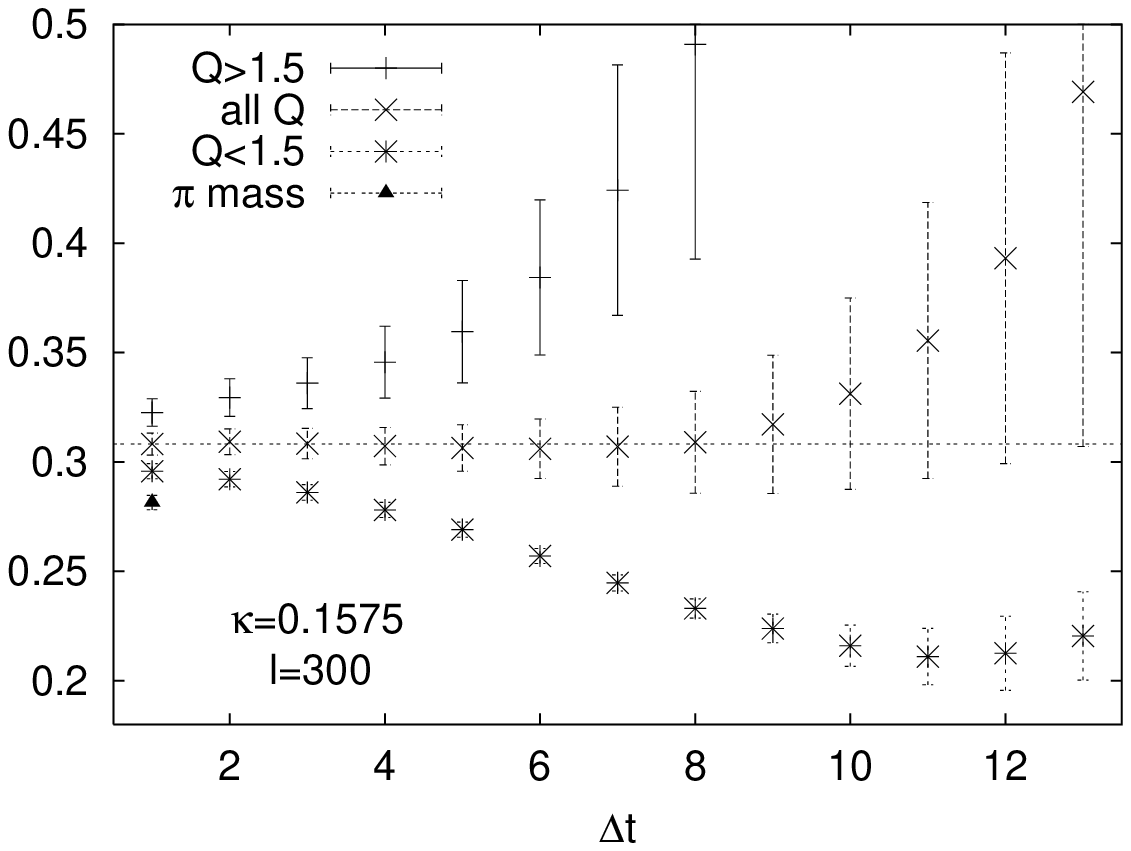,width=1.0\linewidth}} 
\vskip .5cm
\caption{Dependence of the local OLGA $\eta'$ mass on the topological charge
work similarly well  for the lightest and heaviest sea quark mass of SESAM. }
\label{fig:tl8}
\end{figure}

We define the truncated two-loop correlator $T^l$ through 
\begin{eqnarray}\label{eq:tlwg} 
&&T^l(\Delta t) =  \\
&&\sum_t \sum_i^l \frac{1}{\lambda_i}
\frac{ \langle \psi_i(t) | \psi_i(t) \rangle}
{ \langle \psi_i | \psi_i \rangle} 
\sum_j^l \frac{1}{\lambda_j}
\frac{\langle \psi_j(t+\Delta t) | \psi_j(t+\Delta t) \rangle}
{\langle \psi_j | \psi_j \rangle}\; , \nonumber 
\end{eqnarray}
and plot in \fig{fig:tl1} the dependence of $T^{l} (\Delta t)$ on the
cutoff $l$.  Contrary to the pion propagator (see \fig{fig:corr1}),
\fig{fig:tl1} indeed reveals good saturation of the spectral
representation by the low-lying eigenmodes over the entire $\Delta
t$-range.

As a check for consistency we compare the local two-loop correlators
from TEA and standard SET at our lightest quark mass in
\fig{fig:tl2}. They are seen to agree very well within their
errors. Note that the TEA data show a much smoother behavior in
$\Delta t$. Additional smearing for SET diminishes those fluctuations.
We notice that TEA and SET data bear errors of equal size. We view
this as an independent confirmation of previous claims that the errors
on the $\eta'$ mass from state of the art SET analyses are dominated
by gauge field noise~\cite{Struckmann:2000bt,McNeile:2000hf}.

Next we come to the more stringent test: local effective
$\eta'$-masses.  According to \eq{eq:etas}, the $\eta'$-propagator is
the difference of one- and two-loop corrrelators, $C_{\pi}$ and $T$.
The ground state contribution to the former, $C^g_{\pi}$, can be
determined very accurately by the standard methods (by iterative
solvers) known from the octet spectrum~\cite{Eicker:1998sy}.  Hence,
it appears very natural to replace the one-loop correlator by its
ground state component, $C^g_{\pi}(\Delta t)$, see \eq{eq:fit}. In the
following, we perform a `one-loop groundstate analysis' (OLGA) by the
extraction of local masses, $m_{\eta'}(\Delta t)$, from the
combination
\begin{equation}
\label{eq:neffs}
\tilde{C}_{\eta'}(\Delta t)=C^g_{\pi}(\Delta t)- N_f T(\Delta t) \; . 
\end{equation}

The results are presented in \fig{fig:tl3}, both for the lightest and
heaviest sea quark masses of SESAM, with and without smearing.  We
find striking plateau formation from the very first time slice
onwards.

As a consistency check and first test of the synthetic data approach,
\eq{eq:neffs}, we compare the local effective masses from TEA and SET
in \fig{fig:compstoch}.  The data points are seen to agree very well
with each other, the TEA points being slightly less fluctuating. The
horizontal lines in \fig{fig:tl3} and \fig{fig:compstoch} refer to the
fitted plateau values for the $\eta'$ masses.

As yet another test on systematic errors we plot in \fig{fig:tl6} -- again for
local and smeared wave functions and the lightest and heaviest quark masses --
the dependence of the local masses $m_{\eta'}(\Delta t)$ from TEA on the
spectral cutoff $l$. It appears that the systematic errors from this cutoff are
well under control, once we truncate the spectral representation with $l
\simeq 150 $ and higher.  Furthermore the data appear to support the idea that
TEA improves when decreasing the quark mass.

Finally we address the question, to what extent the $\eta'$ mass is
influenced by the topological content of the configurations.  By
applying the cuts in $Q$, we subdivide the gauge field ensemble for
the two $\kappa$-values into two subsets each, with $Q$ determined as
described in section \ref{sec:bechmarking}.  The cuts are chosen such
that each subset consists of about 100 configurations.  The results
from OLGA can be seen in \fig{fig:tl8}.  They clearly confirm the
previous finding~\cite{Struckmann:2000bt,Bali:2001gk}, that
topologically nontrivial gauge configurations are the origin for the
large $\eta'$ mass~\cite{Witten:1979vv,Veneziano:1979ec}.

\section{Discussion and summary}

We have presented and validated a method to compute fermion loops from the
low-lying eigenmodes of the hermitian form, $Q=\gamma_5 M$, of the
standard Dirac-Wilson matrix, $M$, in accordance with the expectation that
these modes contain the essential physics associated with topological
fluctuations.  Our truncated spectral approach (TEA) to $Q^{-1}$ is viable in
the sense that it renders satisfying results in the quark mass regime of
state-of-the art full QCD simulations like SESAM on the basis of ${\cal
  O}(100)$ modes only. TEA has been verified both with respect to the
topological charge and the two-loop correlator entering the $\eta'$
propagator.

The early onset of saturation for $\trq ^{-1}$ could be attained by proper
ordering and subsequent partial summations, configurationwise adapted to
achieve cancellations from positive and negative eigenvalues.  In this way the
bulk of the higher mode contributions were shown to vanish.  For the case of
the $\eta'$ an early plateau formation of the local masses could be obtained
by ground state projecting the connected piece of its propagator.

\begin{figure}[t]
\centering{\epsfig{figure=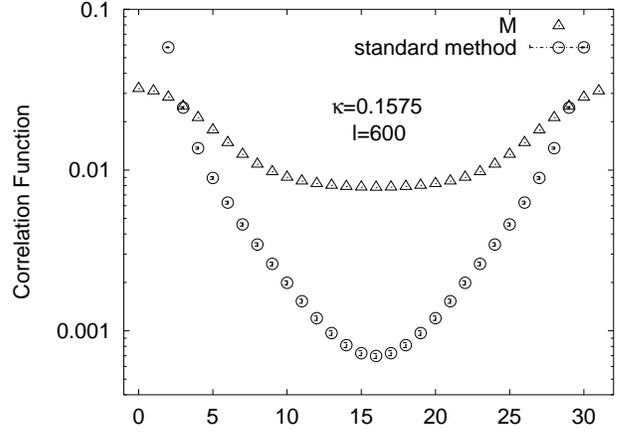,width=1.0\linewidth}}
\vskip .5cm
\caption{Pion correlation function from TEA with $600$ eigenmodes of
$M$ on one configuration for $\kappa=0.1575$. One observes strong
deviations from the standard propagator (from linear solvers)
determined on the entire gauge field ensemble.}
\label{fig:M_CORR}
\end{figure}

\begin{figure}[t]
\centering{\epsfig{figure=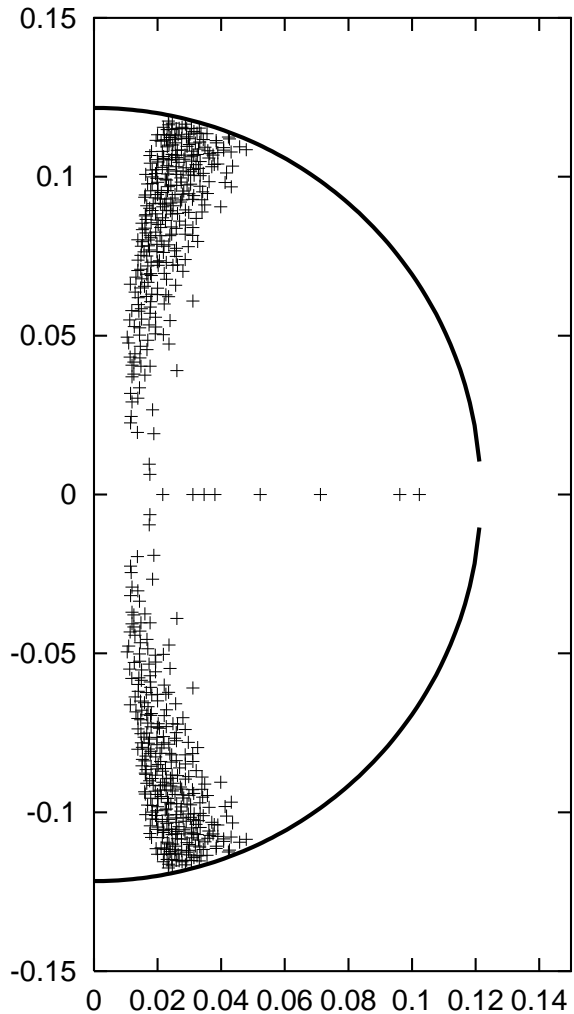,width=2.3\linewidth}}
\vskip .5cm
\caption{The 600 calculated eigenvalues of $M$ that enter
Fig.~\ref{fig:M_CORR}. The curve represents a circle around the origin.}
\label{fig:m_spec}
\end{figure}
\begin{figure}[t]
\centering{\epsfig{figure=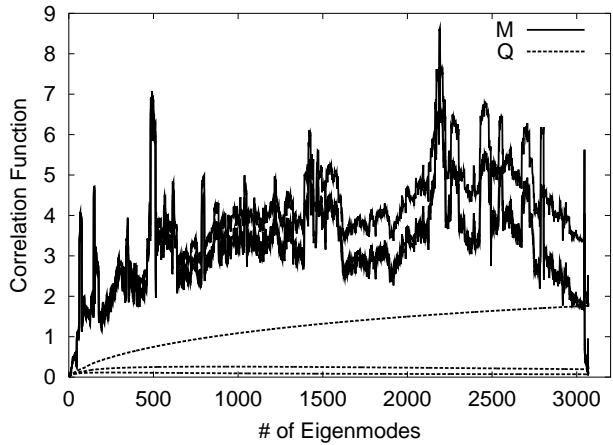,width=1.0\linewidth}} 
\vskip .5cm
\caption{Pion correlation function from TEA for $M$ and $Q$ on a
quenched $4^4$ lattice at $\beta=5.0$. The plot is analogous to
Fig.~\ref{fig:corr1}.}
\label{fig:mcorr}
\end{figure}

We confirm previous results in the intermediate quark mass regime that the
actual bottleneck in increasing accuracy of loop estimates is given by the
gauge field noise, i.e.\ by the present limitation in ensemble sizes of QCD
vacuum configurations.

We found that the amount of work, i.e.\ the number of matrix vector
multiplies, $N^{mvm}$, needed per configuration in TEA compared to
SET~\cite{Struckmann:2000bt} for the lightest SESAM quark mass is
roughly in the same ballpark\footnote{Actually the ratio
$N^{mvm}_{TEA}/N^{mvm}_{SET} \simeq 1.5$ when using 300 stochastic
source vectors and 300 eigenmodes.}.  This looks promising for the
upcoming era of Teracomputing, where we shall deal with larger
lattices and smaller quark masses. The reason is, that the Arnoldi
method does not lose efficiency when entering deeper into the
critical chiral regime, in contrast to Krylov solvers used within
stochastic estimator algorithms which will suffer in convergence rate.

Let us finally comment on the viability of the spectral approach
applied to $M$ instead of $Q$. To this end we compare in
\fig{fig:M_CORR} the pion correlation function as determined from 600
low-lying eigenmodes of $M$, see \fig{fig:m_spec}, with the one
obtained from the standard method (solving linear systems).
Obviously, for the case of $\kappa = 0.1575$ which corresponds to a
pion mass of the order of 730 MeV, the chosen low-lying modes of $M$
contain less information about this function than the eigenmodes of
$Q$. In order to trace down this discrepancy between $M$ and $Q$ we
also carried out a full diagonalization on a quenched $4^4$ lattice at
$\beta=5.0$ for the two cases.  In \fig{fig:mcorr}, we plotted the
resulting $C_{\pi}(\Delta t)$ versus the cutoff.  While $Q$ again
shows a quite stable behavior similar to \fig{fig:corr1}, the spectral
approximation for $M$ yields a very ragged cutoff dependence that
requires all 3072 eigenmodes for a satisfactory representation of the
correlator.  We conclude that the eigenmodes of $M$, being non
orthogonal, suffer interferences among each other.  Thus, for the
$4^4$ test case and for the sea quark masses used in the SESAM
configurations, we have not been able to identify a limited number of
dominating eigenmodes.

A detailed analysis of the $\eta'$ mass, based on OLGA and additional
SET data, will be presented in a forthcoming paper~\cite{neff:2001iq}.

{\bf Acknowledgements} The bulk of computations for this project was
carried out at NIC/J\"ulich and NERSC/Livermore. We thank the staff of
both centers for their support. Complementary analysis was carried out
on the 128 node cluster ALiCE with the support of the DFG project
Li701/3-1.  H.\ N.\ thanks G.\ Bali, I.\ Hip, B.\ Orth, W.\ Schroers,
T.\ Struckmann and P.\ Ueberholz for discussions and support.  The HMC
productions were run on the APE100 systems at INFN Roma and NIC
Zeuthen.  We are grateful to our colleagues G.\ Martinelli and F.\
Rapuano for the fruitful T$\chi$L-collaboration.  Work supported in
part by the U.S. Department of Energy (DOE) under cooperative research
agreement \#DE-FC02-94ER 40818.

\bibliographystyle{h-elsevier}
\bibliography{lit,/home/neff/BIBLIOGRAPHY/00_01,/home/neff/BIBLIOGRAPHY/97_99,/home/neff/BIBLIOGRAPHY/94_96,/home/neff/BIBLIOGRAPHY/proceedings}

\end{document}